\documentclass[hyper,letterpaper,notoc]{JHEP3}

\usepackage{pstricks}
\usepackage{epsfig}
\usepackage{amsbsy}
\usepackage{varioref}
\usepackage{pifont}
\usepackage{amsmath}
\usepackage{graphicx}
\usepackage{axodraw}

\def\slashii#1{\setbox0=\hbox{$#1$}             % set a box for #1
   \dimen0=\wd0                                 % and get its size
   \setbox1=\hbox{\sl/} \dimen1=\wd1            % get size of /
   \ifdim\dimen0>\dimen1                        % #1 is bigger
      \rlap{\hbox to \dimen0{\hfil\sl/\hfil}}   % so center / in box
      #1                                        % and print #1
   \else                                        % / is bigger
      \rlap{\hbox to \dimen1{\hfil$#1$\hfil}}   % so center #1
      \hbox{\sl/}                               % and print /
   \fi}                                         %
\def\slashiii#1{\setbox0=\hbox{$#1$}#1\hskip-\wd0\hbox to\wd0{\hss\sl/\/\hss}}
\title{A Three Site Higgsless Model}

\author{R. Sekhar Chivukula, Baradhwaj Coleppa, Stefano Di Chiara, and Elizabeth H. Simmons\\
Department of Physics and Astronomy, Michigan State University\\
East Lansing, MI 48824, USA\\
E-mail: sekhar@msu.edu, baradhwa@msu.edu, dichiara@msu.edu, esimmons@msu.edu}

\author{Hong-Jian He\\
Center for High Energy Physics, Tsinghua University\\
Beijing 100084, China\\
	E-mail: hjhe@mail.tsinghua.edu.cn}

\author{
Masafumi Kurachi\\
C.N. Yang Institute for Theoretical Physics, State University of New York\\
Stony Brook, NY 11794, USA\\
	E-mail: masafumi.kurachi@stonybrook.edu}

\author{
Masaharu Tanabashi\\
Department of Physics, Tohoku University\\
Sendai 980-8578, Japan\\
	E-mail: tanabash@tuhep.phys.tohoku.ac.jp}

\abstract{
We analyze the spectrum and properties of
a highly-deconstructed Higgsless model with only three sites.  Such a model contains sufficient complexity to incorporate interesting physics issues related to fermion masses and electroweak observables, yet remains simple enough that it could be encoded in a Matrix Element Generator program for use with Monte Carlo simulations.  The gauge sector of this model is equivalent to that of the Breaking Electroweak Symmetry Strongly (BESS) model; the new physics of interest here lies in the fermion sector.  We analyze the form of
the fermion Yukawa couplings required to produce the ideal fermion delocalization that causes
tree-level precision electroweak corrections to vanish.  We discuss the size of one-loop corrections to
$b\to s \gamma$, the weak-isospin violating parameter $\alpha T$ and the decay $Z \to b \bar{b}$.
We find that the new fermiophobic vector states (the analogs of the gauge-boson KK
modes in a continuum model) can be reasonably light, with a mass
as low as 380 GeV, while the extra (approximately vectorial) quark and lepton states 
(the analogs of the fermion KK modes) must be heavier than 1.8 TeV.
\\ \\ 
\centerline{Sept. 7, 2006}}


\preprint{MSUHEP-060710\\
YITP-SB-06-27\\
TU-773}

\begin{document}

%%%%%%%%%%%%%%%%%%%%%%%%%%%%%%%%%%%%
%%%%%%%%%%%%%%%%%%%%%%%%%%%%%%%%%%%%
\section{Introduction}

Higgsless models \cite{Csaki:2003dt} literally 
break the electroweak symmetry without invoking a scalar Higgs  boson 
\cite{Higgs:1964ia}.   Among the most popular are 
 models \cite{Agashe:2003zs,Csaki:2003zu,Burdman:2003ya,Cacciapaglia:2004jz} 
 based on a five-dimensional
$SU(2) \times SU(2) \times U(1)$ gauge theory in a slice of Anti-deSitter space, with
electroweak symmetry breaking  encoded in the boundary conditions of the gauge fields.   The spectrum includes states identified with the photon, $W$, and $Z$, and also  an infinite tower of  
additional massive vector bosons (the higher Kaluza-Klein  or $KK$ excitations) starting at the 
TeV scale \cite{Antoniadis:1990ew}, whose exchange unitarizes longitudinal $W$ and $Z$ boson scattering \cite{SekharChivukula:2001hz,Chivukula:2002ej,Chivukula:2003kq,He:2004zr}. The properties of Higgsless models may be studied  \cite{Foadi:2003xa,Hirn:2004ze,Casalbuoni:2004id,Chivukula:2004pk,Perelstein:2004sc,Georgi:2004iy,SekharChivukula:2004mu} by
using the technique of deconstruction 
\cite{Arkani-Hamed:2001ca,Hill:2000mu} and 
computing the precision electroweak parameters 
\cite{Peskin:1992sw,Altarelli:1990zd,Altarelli:1991fk,Barbieri:2004qk,Chivukula:2004af} in a 
related linear moose model \cite{Georgi:1985hf}. 

Our analysis of the leading electroweak parameters in a very general class of linear moose models concluded \cite{SekharChivukula:2004mu} 
that a Higgsless model with localized fermions cannot simultaneously satisfy unitarity bounds and  provide acceptably small precision electroweak corrections unless it includes light vector bosons other than the photon, $W$, and $Z$.  Several authors proposed \cite{Cacciapaglia:2004rb,Cacciapaglia:2005pa,Foadi:2004ps,Foadi:2005hz} that delocalizing fermions within the extra dimension could reduce  electroweak corrections. In deconstructed language, delocalization means allowing fermions to derive electroweak properties from more than one site on the lattice of gauge groups \cite{Chivukula:2005bn,Casalbuoni:2005rs}. We then showed \cite{SekharChivukula:2005xm} for an arbitrary Higgsless model that choosing the probability distribution of the delocalized fermions to be related to the wavefunction of the $W$ boson makes three ($\hat S$, $\hat T$, $W$) of the leading zero-momentum precision electroweak parameters  defined by Barbieri, et. al. \cite{Barbieri:2004qk,Chivukula:2004af} vanish at tree-level.  We denote such fermions as ``ideally delocalized".   We subsequently provided a continuum realization of ideal delocation that preserves the characteristic of vanishing precision electroweak corrections up to subleading order \cite{SekharChivukula:2005cc}.  In the absence of precision electroweak corrections, the strongest constraints on Higgsless models come from limits on deviations in multi-gauge-boson vertices; we computed the general form of the triple and quartic gauge boson couplings for these models and related them to the parameters of the electroweak chiral Lagrangian \cite{Chivukula:2005ji,Grojean:2006nn}. 

In this paper, we show that many issues of current interest, such as ideal fermion delocalization and the generation of fermion masses (including the top quark mass) can be usefully illustrated in a Higgsless model deconstructed to just three sites.  The Moose describing the model has only one ``interior" $SU(2)$ group and there is, accordingly, only a single triplet of $W'$ and $Z'$ states instead of the infinite tower of triplets present in the continuum limit.  This model contains sufficient complexity to incorporate the interesting physics issues, yet remains simple enough that it could be encoded in a Matrix Element Generator program in concert with a Monte Carlo Event Generator \footnote{ See e.g. those appearing on  http://www-theory.lbl.gov/tools/ .} for a detailed investigation of collider signatures.  

The three-site model we introduce here has the color group of the standard model (SM) and an extended $SU(2) \times SU(2)\times U(1)$ electroweak gauge group.  This theory is in the same class as models of extended  electroweak gauge symmetries \cite{Casalbuoni:1985kq,Casalbuoni:1996qt}  motivated by models of hidden local symmetry \cite{Bando:1985ej,Bando:1985rf,Bando:1988ym,Bando:1988br,Harada:2003jx}.  Indeed the gauge sector is precisely that of the BESS model \cite{Casalbuoni:1985kq}; the new physics discussed here relates to the fermon sector.  In the next section of this paper we briefly describe the model and the limits in which we work.  Section 3 reviews the gauge sector of the model in our notation, including the masses and wavefunctions of the photon, the nearly-standard light $W$ and $Z$ and the heavier $W'$ and $Z'$. Section 4 solves for the masses and wavefunctions of the fermions in the spectrum (a set of SM-like fermions and heavy copies of those fermions) and implements ideal delocalization for the light fermions.   Sections 5 and 6 explore the couplings of the fermions to the charged and neutral gauge bosons, respectively.  Because
the light fermions are ideally delocalized, they lack couplings to the $W'$ and $Z'$ -- and this minimizes the values of electroweak precision observables.  The top quark, on the other hand, is treated separately in order to provide for its large mass.  The relationship of triple gauge vertices to ideal delocalization and a comparison of multi-gauge vertices in the three-site model and its continuum limit are discussed in section 7; given the vanishing electroweak corrections and the fermiophobic nature of the $W'$ and $Z'$, multi-gauge vertices offer the best prospects for additional experimental constraints on the three-site model.   In sections 8 and 9, the paper moves on to  a discussion of $\alpha T$ and the $Zb\bar{b}$ vertex at one loop. Having established that the heavy fermions must have masses of over 1.8 TeV, we discuss the structure of a low-energy effective theory in which those fermions have been integrated out.  Section 10 presents our conclusions.

%%%%%%%%%%%%%%%%%%%%%%%%%%%%%%%%%%%%
%%%%%%%%%%%%%%%%%%%%%%%%%%%%%%%%%%%%
\section{Three Site Model}

\EPSFIGURE[t]{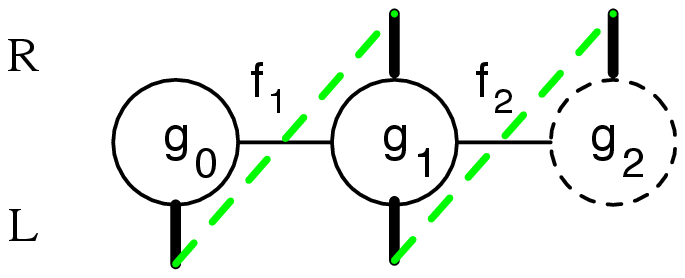,width=0.5\textwidth}
{The three site model analyzed in this paper. The solid circles represent
$SU(2)$ gauge groups, with coupling strengths $g_0$ and $g_1$, and the dashed circle is
a $U(1)$ gauge group with coupling $g_2$. The left-handed fermons,
denoted by the lower vertical lines, are located at sites 0 and 1, and the right-handed fermions,
denoted by the upper vertical lines, at sites 1 and 2. The dashed green lines correspond to
Yukawa couplings, as described in the text. As discussed below, we will take
$f_1 = f_2 =\sqrt{2}\, v$, denote $g_0=g$, $g_1=\tilde{g}$, $g_2=g'$ 
and take $\tilde{g} \gg g, \,g'$.\label{fig:one}}

The electroweak sector of the three-site Higgsless model analyzed in this paper is illustrated in figure \ref{fig:one} using ``Moose notation'' \cite{Georgi:1985hf}.  The model  incorporates an
$SU(2) \times SU(2) \times U(1)$ gauge group, and $2$ 
nonlinear $(SU(2)\times SU(2))/SU(2)$ sigma models in which the global symmetry groups 
in adjacent sigma models are identified with the corresponding factors of the gauge group.
The symmetry breaking between the middle $SU(2)$ and the $U(1)$ follows an $SU(2)_L \times SU(2)_R/SU(2)_V$ symmetry breaking pattern with the $U(1)$ embedded as the $T_3$-generator of $SU(2)_R$.    This extended electroweak gauge sector is that of the Breaking Electroweak Symmetry Strongly (BESS) model \cite{Casalbuoni:1985kq}.

The left-handed fermions are $SU(2)$ doublets coupling to the groups at the first
two sites, and which we will  correspondingly label  $\psi_{L0}$ and $\psi_{L1}$. The
right-handed fermions are an $SU(2)$ doublet at site 1, $\psi_{R1}$, and two singlet
fermions, denoted in figure \ref{fig:one} as ``residing" at site 2, $u_{R2}$ and $d_{R2}$.
The fermions $\psi_{L0}$, $\psi_{L1}$, and $\psi_{R1}$ have $U(1)$ charges typical
of the left-handed doublets in the standard model, $+1/6$ for quarks and $-1/2$ for leptons.
Similarly, the fermion $u_{R2}$ has $U(1)$ charges typical for the right-handed
up-quarks (+2/3), and $d_{R2}$ has the $U(1)$ charge associated with the right-handed
down-quarks ($-1/3$) or the leptons ($-1$). With these assignments, we may write the
Yukawa couplings and fermion mass\footnote{In this paper, we will not address the issue
of nonzero neutrino masses. Our focus, instead, is on the physics related to the generation
of the large top-quark mass.}  term
\begin{equation}
{\cal L}_f = \lambda f_1\, \bar{\psi}_{L0} \Sigma_1 \psi_{R1} + \sqrt{2} \tilde{\lambda} v \,\bar{\psi}_{R1}
\psi_{L1} + f_2 \,\bar{\psi}_{L1} \Sigma_2
\begin{pmatrix}
\lambda'_{u} & \\
& \lambda'_d
\end{pmatrix}
\begin{pmatrix}
u_{R2} \\
d_{R2}
\end{pmatrix}
+ h.c.
\label{eq:yukawa}
\end{equation}
Here we have chosen to write the $\bar{\psi}_{R1} \psi_{L1}$ Dirac mass in the form of
a Yukawa coupling, for convenience, and the matrices $\Sigma_{1,2}$ are the nonlinear
sigma model fields associated with the $f_{1,2}$ links of the moose. The Yukawa couplings
introduced here are of precisely the correct form required to implement a deconstruction
of a five-dimensional fermion with chiral boundary conditions \cite{Hill:2002kq}.
In the limit in which the ``bulk fermion" decouples, while holding the mixing
with the light fermions fixed,  the model reduces to
that considered in \cite{Anichini:1994xx}.

It is straightforward to incorporate quark flavor and mixing in a minimal way. Adding generational indices
to each of the fermion fields, we may choose the coupling $\lambda$ and the mass
term $\sqrt{2} \tilde{\lambda} v$ to be generation-diagonal. In this case, all of the nontrivial
flavor structure is embedded in the Yukawa matrices $\lambda'_u$ and $\lambda'_d$ --
precisely as in the standard model; the only mixing parameters that appear
are the ordinary Cabibbo-Kobayashi-Maskawa (CKM) angles and phase. We focus most of our attention, in this paper, on the
top-bottom quark doublet and their heavy partners, and we note where results for the other fermions differ.

For simplicity, we examine the case
\begin{equation}
f_1 = f_2 = \sqrt{2} v~,
\label{eq:equalfs}
\end{equation}
and work in the limit
\begin{equation}
x={g_0/g_1} \ll 1~,\ \ \ \ \ y={g_2/g_1}\ll 1~,
\end{equation}
in which case we expect a massless photon, light $W$ and $Z$ bosons, and
a heavy set of bosons $W'$ and $Z'$.  Numerically, then, $g_{0,2}$ are approximately
equal to the standard model $SU(2)_W$ and $U(1)_Y$ couplings, and we therefore
denote $g_0 \equiv g$ and $g_2 \equiv g'$, and define an angle $\theta$
such that
\begin{equation}
\frac{g'}{g} = \frac{\sin\theta}{\cos\theta} \equiv \frac{s}{c}~,
\label{eq:scdef}
\end{equation}
In addition, we denote $g_1 \equiv \tilde{g}$.

In what follows, we will show that ideal mixing requires the flavor-independent mass contribution from $\Sigma_1$ to be much smaller than the Dirac mass contribution:
\begin{equation}
\varepsilon_L \equiv \frac{\lambda}{\tilde{\lambda}} = {\cal O}(x) \ll 1~,
\end{equation}
While we will not immediately require that the flavor-dependent mass contributions associated with $\Sigma_2$ 
\begin{equation}
\varepsilon_{uR,dR} \equiv  \frac{\lambda'_{u,d}}{\tilde{\lambda}}~,
\end{equation}
be similarly small, we will ultimately find that they are bounded from above.   The Yukawa and fermion mass terms in the Lagrangian can now be rephrased as
\begin{equation}
{\cal L}_f = \sqrt{2} \tilde{\lambda} v \left[  \varepsilon_L  \, \bar{\psi}_{L0} \Sigma_1 \psi_{R1} +  \bar{\psi}_{R1}
\psi_{L1} + \bar{\psi}_{L1} \Sigma_2
\begin{pmatrix}
\varepsilon_{uR} & \\
& \varepsilon_{dR}
\end{pmatrix}
\begin{pmatrix}
u_{R2} \\
d_{R2}
\end{pmatrix}
+ h.c.\right]
\label{eq:ryukawa}
\end{equation}

\noindent for easy reference.

Finally note that, treating the link fields as non-linear sigma models, the model as described
here is properly considered a low-energy effective theory valid below a mass scale of order
$4 \pi \sqrt{2} v \approx 4.3$ TeV. If we regard each of the link fields as arising from
QCD-like dynamics at that scale, we would expect large corrections to the $S$ parameter
arising from higher-energy operators \cite{Perelstein:2004sc}. On the other hand,
if this model is viewed as the deconstructed form of a five-dimensional
``dual" of some strongly-coupled four-dimensional theory \cite{Maldacena:1998re,Gubser:1998bc,Witten:1998qj,Aharony:1999ti}, the leading corrections are accounted for
by tree-level $W'$-exchange \cite{Burdman:2003ya}. The remaining corrections are suppressed in
the large-$N$ expansion, and may be sufficiently small to be phenomenologically
acceptable. 

%%%%%%%%%%%%%%%%%%%%%%%%%%%%%%%%%%%%
%%%%%%%%%%%%%%%%%%%%%%%%%%%%%%%%%%%%
\section{Masses and Eigenstates}

This section reviews the mass eigenvalues and the wavefunctions of the
gauge bosons  of the three-site model, which are the same as those for the BESS model \cite{Casalbuoni:1985kq}.  Ref. \protect\cite{Foadi:2003xa} has also previously discussed the gauge boson eigenfunctions, but wrote them in terms of the parameters e, $M_W$, $M_Z$, $M_{W'}$, and $M_{Z'}$.
 
%%%%%%%%%%%%%%%%
%%%%%%%%%%%%%%%%
\subsection{Charged Gauge Bosons}

The charged gauge-boson mass-squared matrix may be written in terms of the small parameter $x$ as 
\begin{equation}
{\tilde{g}^2 v^2 \over 2}
\begin{pmatrix}
x^2 & -x \\
-x & 2 
\end{pmatrix}~.
\end{equation}
Diagonalizing this matrix perturbatively in $x$, we find the light eigenvalue
\begin{equation}
M^2_W = {g^2 v^2 \over 4} \left[
1-{x^2\over 4}+{x^6 \over 64} + \ldots
\right]~,
\label{eq:mwsq}
\end{equation}
and the corresponding eigenstate
\begin{align}
W^\mu  & = v^0_W\, W^\mu_0 + v^1_W \, W^\mu_1 \nonumber\\
& = \left(1-{x^2 \over 8} - {5 x^4 \over 128} + \ldots \right) W^\mu_0
+
\left({x\over 2} +{x^3\over 16}-{9x^5 \over 256}+\ldots \right) W^\mu_1~,
\label{eq:weigenvector}
\end{align}
where $W_{0,1}$ are the gauge bosons associated with the $SU(2)$ groups
at sites 0 and 1.   Note that the light $W$ is primarily located at site 0.

The heavy eigenstate has an eigenvector orthogonal to that in eqn. (\ref{eq:weigenvector}) and a mass
\begin{equation}
M^2_{W'}=\tilde{g}^2 v^2 \left[ 1+{x^2\over 4}+{x^4\over 16} + \ldots \right]~,
\label{eq:mwheavysq}
\end{equation}
Comparing eqns. (\ref{eq:mwsq}) and (\ref{eq:mwheavysq}), we find
\begin{equation}
{M^2_W \over M^2_{W'}} = {x^2\over 4} - {x^4 \over 8} + {x^6\over 64} + \ldots~,
\label{eq:mwratio}
\end{equation}
or, equivalently,
\begin{equation}
\left({g_0\over g_1}\right)^2 \equiv x^2 = 4\left({M^2_W \over M^2_{W'}}\right) + 8 \left({M^2_W \over M^2_{W'}}\right)^2
+28\left({M^2_W \over M^2_{W'}}\right)^3+\ldots\, ,
\label{eq:xequation}
\end{equation}
which confirms that the $W'$ boson is heavy in the limit of small $x$.

%%%%%%%%%%%%%%
%%%%%%%%%%%%%%
\subsection{Neutral Gauge Bosons}

The neutral bosons' mass-squared matrix is
\begin{equation}
{\tilde{g}^2 v^2 \over 2}
\begin{pmatrix}
x^2 & -x & 0 \\
-x & 2 & -x t \\
0 & -x t & x^2 t^2
\end{pmatrix}~,
\end{equation}
where $t \equiv \tan\theta = s/c$. This matrix has a zero eigenvalue, corresponding
to the massless photon, with an eigenstate which may be written
\begin{equation}
A^\mu = {e\over g} W^\mu_0 + {e \over \tilde{g}} W^\mu_1 + {e\over g'} B^\mu~,
\end{equation}
where $W_{0,1}$ are the gauge bosons associated with the $SU(2)$ groups at
sites 0 and 1, the $B$ is the gauge boson associated with the $U(1)$ group at site
2, and the electric charge $e$ satisfies
\begin{equation}
{1\over e^2} = {1\over g^2} + {1\over \tilde{g}^2} + {1\over {g'}^2}~.
\label{eq:edef}
\end{equation}

The light neutral gauge boson, which we associate with the $Z$, has a mass
\begin{equation}
M^2_Z = {g^2 v^2 \over 4\, c^2} \left[
1-{x^2\over 4}{(c^2-s^2)^2 \over c^2}+{x^6\over 64} {(c^2-s^2)^4 \over c^6} + \ldots
\right]~,
\label{eq:zmasssq}
\end{equation}
with a corresponding eigenvector
\begin{align}
Z^\mu = v^0_Z W^\mu_0 &+ v^1_Z W^\mu_1 + v^2_Z B^\mu \\
v^0_Z & = c-{x^2 c^3(1+2t^2-3 t^4)\over 8}+\ldots\\
v^1_Z & =  {x c (1-t^2)\over 2}+{x^3 c^3 (1-t^2)^3 \over 16} + \ldots \\
v^2_Z & = -s -{x^2s c^2(3-2t^2-t^4) \over 8}+\ldots~.
\end{align}
The heavy neutral boson has a mass
\begin{equation}
M^2_{Z'} = \tilde{g}^2 v^2 \left[1+{x^2 \over 4 c^2}+{x^4(1-t^2)^2 \over 16} + \ldots\right]~,
\end{equation}
with the corresponding eigenvector
\begin{align}
{Z'}^\mu = v^0_{Z'} W^\mu_0 &+ v^1_{Z'} W^\mu_1 + v^2_{Z'} B^\mu \\
v^0_{Z'} & = -{x\over 2} -{x^3 (1-3t^2)  \over 16}+\ldots \\
v^1_{Z'} & = 1-{x^2(1+t^2) \over 8} + \ldots \\
v^2_{Z'} & = -{xt\over 2} + { x^3 t (3-t^2) \over 16} + \ldots~.
\end{align}
%

%%%%%%%%%%%%%%%%%%%%%%%%%%%%%%%%%%%%
%%%%%%%%%%%%%%%%%%%%%%%%%%%%%%%%%%%%
\section{Fermion Wavefunctions and Ideal Delocalization}

This section analyzes the fermion sector of the three-site model and implements ideal fermion delocalization explicitly.

%%%%%%%%%%%%%%
%%%%%%%%%%%%%%
\subsection{Fermion masses and wavefunctions}

Consider the fermion mass matrix
\begin{equation}
M_{u,d} = \sqrt{2} \tilde{\lambda} v
\begin{pmatrix}
\varepsilon_L & 0 \\
1 & \varepsilon_{uR,dR}
\end{pmatrix}  \equiv  \begin{pmatrix}
m & 0 \\
M & m'_{u,d} 
\end{pmatrix}~.
\label{eq:fermmat}
\end{equation}
The notation introduced at the far right is used to emphasize the
``see-saw" form of the mass matrix. In what follows, we will largely be interested
in the top- and bottom-quarks, and therefore in $\varepsilon_{tR}$ and $\varepsilon_{bR}$
(or, equivalently, in $m'_t/M$ and $m'_b/M$).   

Diagonalizing the top-quark seesaw-style mass matrix perturbatively in $\varepsilon_L$, 
we find the light eigenvalue
\begin{align}
m_t & = {\sqrt{2} \tilde{\lambda} v \varepsilon_L \varepsilon_{tR} \over \sqrt{1+ \varepsilon^2_{tR}}}
\left[
1 - {\varepsilon^2_L \over 2\,(\varepsilon^2_{tR}+1)^2}+\ldots
\right]~,\label{eq:mtone}\\
& \approx {m\, m'_t \over \sqrt{M^2+{m'_t}^2}}~.
\label{eq:mt}
\end{align}
Note that this is precisely the same form as found in \cite{Foadi:2005hz}.
For the bottom-quark, we find the same expression with $\varepsilon_{tR} \to 
\varepsilon_{bR}$, and therefore (neglecting higher order terms in $\varepsilon^2_{bR}$)
\begin{equation}
{m_b \over m_t} \approx {\varepsilon_{bR} \over \varepsilon_{tR}} \sqrt{1+\varepsilon^2_{tR}}
\label{eq:massratio}
\end{equation}
The heavy eigenstate ($T$) corresponding to the top-quark has a mass
\begin{align}
m_T & = \sqrt{2} \tilde{\lambda} v \sqrt{1+\varepsilon^2_{tR}}
\left[
1+ {\varepsilon^2_L \over 2\,(\varepsilon^2_{tR}+1)^2}+\ldots
\right]~,\\
& \approx \sqrt{M^2+{m'_t}^2}
\label{eq:massheavyT}
\end{align}
and similarly for the heavy eigenstate corresponding to the bottom-quark ($B$) with 
$\varepsilon_{tR} \to  \varepsilon_{bR}$ (or, equivalently, $m'_t \to m'_b$).

The left- and right-handed light mass eigenstates of the top quark are
\begin{align}
t_L & = t^0_L\, \psi^t_{L0} + t^1_L\,\psi^t_{L1} \nonumber \\
& = \left(-1 + {\varepsilon^2_L \over 2(1+\varepsilon^2_{tR})^2} + {(8\varepsilon^2_{tR}-3)\varepsilon^4_L \over 8(\varepsilon^2_{tR}+1)^4}+\ldots \right) \psi^t_{L0}
+\left({\varepsilon_L \over 1+\varepsilon^2_{tR}} + 
{(2\varepsilon^2_{tR}-1)\varepsilon^3_L\over 2(\varepsilon^2_{tR} + 1)^3} + \ldots\right) \psi^t_{L1} 
\label{eq:fleigenvector}\\
t_R & = t^1_R\, \psi^t_{R1} + t^2_R\,t_{R2} \nonumber \\
& = \left(-\,{\varepsilon_{tR} \over \sqrt{1+\varepsilon^2_{tR}}} + 
{\varepsilon_{tR}\, \varepsilon^2_L \over (1+\varepsilon^2_{tR})^{5/2}}+\ldots\right)\psi^t_{R1}
+\left({1\over \sqrt{1+\varepsilon^2_{tR}}}+{\varepsilon^2_{tR}\,\varepsilon^2_L \over
(1+\varepsilon^2_{tR})^{5/2}}+\ldots\right)t_{R2}~,
\label{eq:freigenvector}
\end{align}
and similarly for the left- and right-handed $b$-quarks with $\varepsilon_{tR} \to 
\varepsilon_{bR}$. Here we denote the upper components of the $SU(2)$ doublet
fields as $\psi^t_{L0,L1,R1}$; clearly the smaller the value of $\varepsilon_L$ ($\varepsilon_{tR}$), the more strongly the left-handed (right-handed) eigenstate will be concentrated at site 0 (site 2). Note that the relative phase of the eigenvectors
$t_L$ and $t_R$ is set by the eigenstate condition
\begin{equation}
M_t^\dagger |t_L\rangle = m_t |t_R\rangle~.
\end{equation}

The left- and right-handed heavy fermion mass eigenstates are the orthogonal combinations
\begin{align}
T_L & = T^0_L \psi^t_{L0} + T^1_L \psi^t_{L1}\\
& = \left(-\,{\varepsilon_L \over 1+\varepsilon^2_{tR}} - 
{(2\varepsilon^2_{tR}-1)\varepsilon^3_L\over 2(\varepsilon^2_{tR} + 1)^3} + \ldots\right) \psi^t_{L0} 
+ \left(-1 + {\varepsilon^2_L \over 2(1+\varepsilon^2_{tR})^2} + {(8\varepsilon^2_{tR}-3)\varepsilon^4_L \over 8(\varepsilon^2_{tR}+1)^4}+\ldots \right) \psi^t_{L1}\\
T_R & = T^1_R \psi^t_{R1} + T^2_R t_{R2}~,\\
&= \left(-\,{1\over \sqrt{1+\varepsilon^2_{tR}}}-{\varepsilon^2_{tR}\,\varepsilon^2_L \over
(1+\varepsilon^2_{tR})^{5/2}}+\ldots\right)\psi^t_{R1} +
\left(-\,{\varepsilon_{tR} \over \sqrt{1+\varepsilon^2_{tR}}} + 
{\varepsilon_{tR}\, \varepsilon^2_L \over (1+\varepsilon^2_{tR})^{5/2}}+\ldots\right)t_{R2}~,
\end{align}
and similarly for the left- and right-handed heavy $B$ quarks with $\varepsilon_{tR} \to \varepsilon_{bR}$.

Analogous results follow for the other ordinary fermions and their heavy partners, with the appropriate $\varepsilon_{fR}$ substituted for $\varepsilon_{tR}$ in the expressions above.

%%%%%%%%%%%%%%%%
%%%%%%%%%%%%%%%%
\subsection{Ideal Delocalization}

As shown in \cite{SekharChivukula:2005xm} it is possible to  minimize precision electroweak corrections due to the light fermions by appropriate (``ideal") delocalization of the light fermions along the moose.  Essentially, if we recall that the $W$ is orthogonal to its own heavy KK modes (the $W'$ in the three-site model), then it is clear that relating the fermion profile along the moose appropriately to the $W$ profile can ensure that the $W'$ will be unable to couple to the fermions.  Specifically, at site $i$ we require the couplings and wavefunctions of the ideally delocalized fermion and the $W$ boson to be related as
\begin{equation}
g_i (\psi^f_i)^2 = g_W v^i_W
\end{equation}
In the three-site model, if we write the wavefunction of a delocalized left-handed fermion  $f_L = f^0_L \psi^f_{L0} + f^1_L \psi^f_{L1}$  then ideal delocalization imposes the following condition(having taken the ratio of the separate constraints for $i=0$ and $i=1$):
\begin{equation}
{g (f^0_L)^2 \over \tilde{g} (f^1_L)^2} = {v^0_W \over v^1_W}~.
\label{eq:idealdlc}
\end{equation}

Based on our general expressions for fermion mass eigenstates (eqns. (\ref{eq:fleigenvector}) and (\ref{eq:freigenvector})) and the $W$ mass eigenstate (\ref{eq:weigenvector}), it is clear that 
(\ref{eq:idealdlc}) relates the flavor-independent quantities $x$ and $\varepsilon_L$ to the flavor-specific $\epsilon_{fR}$.  Hence, if we construe this as an equation for $\varepsilon_L$ and solve perturbatively in the small quantity $x$, we find\footnote{In the three-site model, this choice of $\varepsilon^2_L$ is equivalent to a choice of
the parameter $b$ in \cite{Anichini:1994xx} to make $\epsilon_3$ or $\alpha S$ vanish.
}   
\begin{equation}
\varepsilon^2_{L} 
\to (1+\varepsilon^2_{fR})^2 \left[
{x^2 \over 2} + \left({1\over 8} - {\varepsilon^2_{fR} \over 2}\right) x^4 +{5\,\varepsilon^4_{fR}\, x^6 \over 8}
+ \ldots
\right]~.
\label{eq:idealf}
\end{equation}
Regardless of the precise value of $\varepsilon_{fR}$ involved, it is immediately clear that ideal delocaliztion implies $\varepsilon_L ={\cal O}(x)$.  Since $x \ll 1$, this justifies the expansions used above in diagonalizing the fermion mass matrix.

The value of $\varepsilon_L$ that yields precisely ideal delocalization for a given fermion species depends on $\varepsilon_{fR}$ and therefore (\ref{eq:massratio}) on the fermion's mass.  For example, the value of $\varepsilon_L$ that ideally delocalizes the $b$ depends on $\varepsilon_{bR}$. As we will see below, however, bounds on the right-handed $Wt b$ coupling will yield the constraint $\varepsilon_{bR} \le 1.4 \times 10^{-2}$; when eqn. (\ref{eq:idealf}) is applied to the $b$ quark and this constraint is imposed, terms proportional to $\varepsilon_{bR}$ become negligible.  As all other fermions (except top) are even lighter, the associated values of $\varepsilon_{fR}$ will be even smaller.  In practice, therefore, 
we may neglect all terms proportional to $\varepsilon_{fR}$
in eqn. (\ref{eq:idealf}), and the condition for ideal mixing is essentially the same for
all fermions except the top-quark: 
\begin{equation}
\varepsilon^2_L = {x^2\over 2} +{x^4 \over 8} + {\cal O}(x^8) =
2\left({M^2_W \over M^2_{W'}}\right) + 6 \left({M^2_W \over M^2_{W'}}\right)^2
+22 \left({M^2_W \over M^2_{W'}}\right)^3+\ldots~,
\label{eq:ideal}
\end{equation}
where the second equality follows from eqn. (\ref{eq:xequation}).
This is the value of $\varepsilon_L$ we will henceforth use for all fermions in our analysis.
 As discussed in \cite{Chivukula:2005ji},  we expect that
the value of $x$ will be bounded by constraints on the $WWZ$ vertex when the light fermions are ideally delocalized.

%%%%%%%%%%%%%%%%%%%%%%%%%%%%%%%%%%%%
%%%%%%%%%%%%%%%%%%%%%%%%%%%%%%%%%%%%
\section{Fermion couplings to the $W$ boson}

%%%%%%%%%%%%%%%%
%%%%%%%%%%%%%%%%
\subsection{Left-handed fermion couplings to the $W$ boson}

We may now compute the couplings between left-handed fermions (ordinary or heavy partners) and the light $W$ boson\footnote{See also the BESS results \cite{Casalbuoni:1985kq}.}. In terms of the mass-eigenstate gauge fields, the left-handed couplings of the
light $W$'s may be written
\begin{equation}
{\cal L}_{WL} \propto W^+_\mu \left[g\, v^0_W(\bar{\psi}_{L0} \tau^- \gamma^\mu \psi_{L0})
+ \tilde{g}\, v^1_W(\bar{\psi}_{L1} \tau^- \gamma^\mu \psi_{L1})\right]  + h.c.~.
\end{equation}
The couplings of the light-$W$ to the mass-eigenstate fermions is then computed
by decomposing the gauge-eigenstate fermions into mass-eigenstates.

We begin with the left-handed $Wtb$ coupling,
assuming ideal mixing for the $b$-quark in the $\varepsilon_{bR} \to 0$ limit.  Because the $W$ wavefunction receives contributions from sites 0 and 1 only, the $Wff'$ coupling is the sum of the overlap between the $W$ and fermion wavefunctions on those two sites:
\begin{equation}
g^{Wtb}_L=g t^0_L b^0_L v^0_W + \tilde{g} t^1_L b^1_L v^1_W~;
\end{equation}
we find
\begin{equation}
g^{Wtb}_L = g \left(1-{3 \varepsilon^4_{tR} + 4 \varepsilon^2_{tR}+3 \over 
8(\varepsilon^2_{tR}+1)^2}\,x^2 + {3 \varepsilon^8_{tR}+16 \varepsilon^6_{tR}+50 \varepsilon^4_{tR}
+ 8\varepsilon^2_{tR} + 15 \over 128(\varepsilon^2_{tR}+1)^4}\, x^4 + \ldots \right)~.
\label{eq:gWtb}
\end{equation}
The corresponding equation for the coupling of standard model fermions other than the top-quark to the $W$ may be obtained by taking $\varepsilon_{tR} \to 0$ in the equation above, yielding
\begin{equation}
g^W_L = g \left( 1-{3 \over 8}\, x^2 + {15 \over 128} x^4 + \ldots\right)~.
\label{eq:gW}
\end{equation}
Combining this  with eqns. (\ref{eq:scdef}), (\ref{eq:mwsq}), (\ref{eq:edef}),  and (\ref{eq:zmasssq}) we find
\begin{equation}
g^W_L = {e \over \sqrt{1-{M^2_W \over M^2_Z}}}
\left[1+{\cal O}(s^2\,x^4)\right]~,
\end{equation}
which shows that the $W$-fermion couplings (for fermions other than top) are of very nearly
standard model form, as consistent with ideal delocalization.   Eqn. (\ref{eq:gW}) corresponds to a value of $G_F$
\begin{equation}
\sqrt{2} G_F = {(g^W_L)^2 \over 4 M^2_W} = {1\over v^2}
\left(1-{x^2\over 2} + {x^4 \over 4} + \ldots\right)~,
\label{eq:GF}
\end{equation}
and the relation
\begin{equation}
g^{Wtb}_L = g^W_L \left(1+{\varepsilon^2_{tR}   \over 4(\varepsilon^2_{tR}+1)^2} x^2
-{\varepsilon^2_{tR}(3 \varepsilon^6_{tR} + 8 \varepsilon^4_{tR}+4 \varepsilon^2_{tR}+10)
\over 32 (\varepsilon^2_{tR}+1)^4} x^4+\ldots\right)~.
\end{equation}

The $W$ also couples to the heavy partners of the ordinary fermions.  Here, we quote the results for the $T$ and $B$ fermions; analogous results follow for other generations when $\varepsilon_{tR}$ is replaced by the appropriate $\varepsilon_{qR}$.  There is a diagonal $WTB$ coupling of the form
\begin{align}
g^{WTB}_L & =  g T^0_L B^0_L v^0_W + \tilde{g} T^1_L B^1_L v^1_W~,\\
& = {g \over 2} \left(1-{\varepsilon^4_{tR}-6 \varepsilon^2_{tR}-5 \over 8(\varepsilon^2_{tR}+1)^2}\,
x^2 + \ldots\right) \\
& = {g^W_L \over 2} \left(1+{\varepsilon^4_{tR} + 6 \varepsilon^2_{tR} + 4 \over 
4 (\varepsilon^2_{tR}+1)^2} x^2  + \ldots\right)~,
\label{eq:gWTB}
\end{align}
where $T^{0,1}_L$ and $B^{0,1}_L$ are the heavy-fermion analogs of the 
components $t^{0,1}_L$ and $b^{0,1}_L$. There are also smaller off-diagonal couplings involving one heavy and one ordinary fermion
\begin{align}
g^{WTb}_L & =  g T^0_L b^0_L v^0_W + \tilde{g} T^1_L b^1_L v^1_W~,\\
& = {g (1-\varepsilon^2_{tR})  \over 2 \sqrt{2}(\varepsilon^2_{tR}+1)} \left(x 
+{\cal O}(x^3)\right)~,
\label{eq:gWTb}
\end{align}
and
\begin{align}
g^{WtB}_L & =  g t^0_L B^0_L v^0_W + \tilde{g} t^1_L B^1_L v^1_W~,\\
& = {g(1+2\varepsilon^2_{tR}) \over 2 \sqrt{2}(\varepsilon^2_{tR}+1)}\left(x 
+{\cal O}(x^3)\right)~,
\label{eq:gWtB}
\end{align}
which play an important role in radiative corrections.

%%%%%%%%%%%%%%
%%%%%%%%%%%%%%
\subsection{Weak mixing angle}

From Eqs.~(\ref{eq:edef}), (\ref{eq:zmasssq}) and (\ref{eq:GF}) 
we can calculate the``$Z$ standard" 
weak mixing angle\footnote{See also the BESS results \cite{Casalbuoni:1985kq,Anichini:1994xx}.} $\theta_W\vert_Z$:
\begin{eqnarray}
 s_Z^2 c_Z^2 &\equiv& \frac{e^2}{4\sqrt{2} G_F M_Z^2} \nonumber \\
 &=& s^2c^2 + s^2(c^2-s^2)\left(c^2-\frac{1}{4}\right)x^2 
 + O(x^4),
\end{eqnarray}
where $s_Z \equiv \sin \theta_W\vert_Z$ and 
$c_Z \equiv \cos \theta_W\vert_Z$.
The relationship between the weak mixing angle $\theta_W\vert_Z$ and 
the angle $\theta$ defined in Eq.~(\ref{eq:scdef}) is expressed 
as follows:
\begin{equation}
 s_Z^2 = s^2 + \Delta,\ \ \ c_Z^2 = c^2 - \Delta,
\label{eq:Zstandard}
\end{equation}
\begin{equation}
 \Delta \equiv s^2 \left(c^2-\frac{1}{4}\right)x^2 + O(x^4).
\label{eq:Delta}
\end{equation}
In other words, $s^2$ and $s^2_Z$ differ by corrections of order $x^2$.

%%%%%%%%%%%%%%%%
%%%%%%%%%%%%%%%%
\subsection{Right-handed fermion couplings to the $W$ boson and $b \to s \gamma$}

Because $\psi_R$ is a doublet under $SU(2)_1$, the three-site model includes  a right-handed
couplings of the $W$
\begin{equation}
{\cal L}_{WR} \propto W^+_\mu  \left[ \tilde{g}\, v^1_W(\bar{\psi}_{R1} \tau^- \gamma^\mu \psi_{R1})\right]  + h.c.~.
\end{equation}
Note that the right-handed fermions exist only on sites 1 and 2 while the $W$ is limited to sites 0 and 1; hence, the right-handed  coupling comes entirely from the overlap at site 1. 
For the $tb$ doublet we find
\begin{align}
g^{Wtb}_R & = \tilde{g}\,t^1_R\,b^1_R\,v^1_W \\
& = {g \over 2} {\varepsilon_{tR}\over \sqrt{1+\varepsilon^2_{tR}}} {\varepsilon_{bR}\over \sqrt{1+\varepsilon^2_{bR}}}\left(1+{\cal O}(x^2)\right)\label{eq:gWRtbqq}\\
& \approx {g\over 2} {m_b \over m_t}{\varepsilon^2_{tR} \over 1+\varepsilon^2_{tR}}~~,
\label{eq:gWRtb}
\end{align}
where reaching the last line requires use of eqn. (\ref{eq:massratio}). It is
interesting to note that this expression is precisely analogous to the related expression
in the continuum model (see eqn. (4.17) of  \cite{Foadi:2005hz}).   

The right-handed $Wtb$ coupling can yield potentially large contributions to $b \to s \gamma$.
As shown in \cite{Larios:1999au}, agreement with the experimental upper limit on this process requires 
\begin{equation}
{g^{Wtb}_R \over g^W_L} < 4 \times 10^{-3}~.
\end{equation}
Combining this bound with our expressions for $g^W_L$ (\ref{eq:gW}) and $g^{Wtb}_R$  (\ref{eq:gWRtb}), recalling $x \ll 1$, and using $m_t =175$ GeV, $m_b=4.5$ GeV, yields the constraint
\begin{equation}
\varepsilon_{tR} < 0.67~.
\label{eq:bsgammaconstraint}
\end{equation}
As we shall see below, this constraint will automatically be satisfied for $M>1.8$ TeV --  a mass limit that will be shown to be required for consistency with top-quark mass generation and limits on $\varepsilon_L$.  Finally, combining eqns. (\ref{eq:bsgammaconstraint}) and (\ref{eq:massratio}),
reveals that
\begin{equation}
\varepsilon_{bR} < 1.4 \times 10^{-2}~,
\label{eq:epsilonbRlimit}
\end{equation}
as referred to earlier.  Again, this confirms that the same value of $\varepsilon_L$ can produce nearly perfect ideal delocalization for the $b$ and all of the lighter fermions.

The $W$ also has right-handed couplings to $T$ and $B$, for which we compute the diagonal coupling
\begin{align}
g^{WTB}_R =&  \tilde{g} T^1_R B^1_R v^1_W\\
=& {g \over 2 \sqrt{1+\varepsilon^2_{tR}}}\left(1+{\varepsilon^4_{tR}+6\varepsilon^2_{tR}+1
\over 8(\varepsilon^2_{tR}+1)^2}x^2+\ldots\right) \\
=& {g^W_L \over 2 \sqrt{1+\varepsilon^2_{tR}}}\left(1+{\varepsilon^4_{tR}+3\varepsilon^2_{tR}+1
\over 2(\varepsilon^2_{tR}+1)^2}x^2+\ldots\right)~,
\label{eq:gRWTB}
\end{align}
and the off-diagonal coupling
\begin{align}
g^{WtB}_R = & \tilde{g} t^1_R B^1_R v^1_W \\
=& {g\,\varepsilon_{tR} \over 2 \sqrt{1+\varepsilon^2_{tR}}}\left(1+
{\varepsilon^4_{tR}+2\varepsilon^2_{tR}-3 \over 8 (\varepsilon^2_{tR}+1)^2} x^2 +\ldots\right)\\
=& {g^W_L\,\varepsilon_{tR} \over 2 \sqrt{1+\varepsilon^2_{tR}}}\left(1+
{\varepsilon^2_{tR}(\varepsilon^2_{tR}+2) \over 2 (\varepsilon^2_{tR}+1)^2} x^2 +\ldots\right) ~.
\end{align}
As in the case of $g^{Wtb}_R$, the right-handed coupling $g^{WTb}_R$
turns out to be proportional to $\varepsilon_{bR}$, and is therefore very small.

Other right-handed $Wqq'$ couplings involving the light standard fermions are straightforward to deduce from eqn. ({\ref{eq:gWRtbqq}) and clearly suppressed by the small values of $\epsilon_{qR}$.  Similarly, the off-diagonal $g^{WQq'}_{R}$  are proportional to small $\epsilon_{qR}$.  The diagonal $g^{WQQ'}_R$ are analogous in form to (\ref{eq:gRWTB}).

%%%%%%%%%%%%%%%%%%%%%%%%%%%%%%%%%%%%
%%%%%%%%%%%%%%%%%%%%%%%%%%%%%%%%%%%%
\section{Fermion couplings to the $Z$ Boson}

The $Z$ coupling to fermions may now be computed.  Like the $W$, the $Z$ may couple to a pair of ordinary or heavy-partner fermions, or to a mixed pair with one ordinary and one heavy-partner fermion.
The left-handed coupling of the light $Z$-boson to quark fields may be written
\begin{equation}
{\cal L}_{ZL} \propto
Z_\mu\left[ g\,v_Z^0 (\bar{\psi}_{L0} {\tau^3\over 2} \gamma^\mu  \psi_{L0}) + 
\tilde{g}\,v_Z^1 (\bar{\psi}_{L1} {\tau^3\over 2} \gamma^\mu \psi_{L1})+
{g'\over 6}\,v_Z^2 \left(\bar{\psi}_{L0} \gamma^\mu \psi_{L0}
+\bar{\psi}_{L1} \gamma^\mu \psi_{L1}\right)\right]~,
\label{eq:leftZlag}
\end{equation}
where the first two terms give rise to the left-handed ``$T_3$" coupling and
the last term (proportional to $g'$) gives rise to the left-handed hypercharge
coupling. The expression for leptons would be similar, replacing hypercharge
$1/6$ with $-1/2$.

Similarly, the right-handed coupling of the $Z$ to quark fields is
\begin{equation}
{\cal L}_{ZR} \propto
Z_\mu\left[ \tilde{g}\,v_Z^1 (\bar{\psi}_{R1} {\tau^3\over 2} \gamma^\mu  \psi_{R1}) + 
{g'\over 6}\,v_Z^2 (\bar{\psi}_{R1} \gamma^\mu \psi_{R1})+
g'\,v_Z^2 \left({2\over 3}\bar{u}_{R2} \gamma^\mu u_{R2}
-{1\over 3} \bar{d}_{R2} \gamma^\mu d_{R2}\right)\right]~,
\end{equation}
where the last three terms arise from the hypercharge. For leptons, $1/6 \to -1/2$
in the second term, $2/3 \to 0$ in the third term (for neutrinos), and $-1/3 \to -1$ in
the last term for the charged leptons. For quarks, this expression may
be more conveniently rewritten as
\begin{equation}
{\cal L}_{ZR} \propto
Z_\mu\left[ (\tilde{g}\,v_Z^1-g' v_Z^2) (\bar{\psi}_{R1} {\tau^3\over 2} \gamma^\mu  \psi_{R1}) + 
g'\,v_Z^2 \left({2\over 3}\sum_{i=1,2} \bar{u}_{Ri}\gamma^\mu u_{Ri}
-{1\over 3} \sum_{i=1,2} \bar{d}_{Ri} \gamma^\mu d_{Ri}\right)\right]~,
\label{eq:rightZlag}
\end{equation}
where the last two terms yield the $Z$-couplings to the conventionally-defined right-handed
hypercharge of the quarks, while the first can give rise to a new right-handed ``$T_3$" coupling.

\subsection{Light Fermion couplings to the $Z$ boson}

We now use eqns. (\ref{eq:leftZlag}) and (\ref{eq:rightZlag}) to compute
the couplings of the $Z$ to light fermions. For an ideally localized light fermion $f$,
we find the left-handed coupling to $T_3$ by summing the overlaps of the $Z$ and fermion wavefunctions on sites 0 and 1 (the loci of the $T_3$ charges):
\begin{align}
g^{Zqq}_{3L} & = g (f^0_L)^2 v^0_Z + \tilde{g} (f^1_L)^2 v^1_Z \\
& = g \,c  \left(1 - {x^2 c^2 (3+6t^2-t^4) \over 8} +\ldots\right) \\
& = {e M_W \over M_Z \sqrt{1-{M^2_W \over M^2_Z}}}
\left[1+{\cal O}(s^2\,x^4)\right]~.
\label{eq:smi}
\end{align}

The coupling of left-handed light fermions to hypercharge arises from the overlap between the fraction of the $Z$ wavefunction arising from site 2 (the locus of hypercharge) and the left-handed fermion wavefunctions which are limited to sites 0 and 1:
\begin{align}
g^{Zqq}_{YL} & = g' v^2_Z \left[(f^0_L)^2 + (f^1_L)^2\right] = g' v^2_Z \label{eq:smiix}\\
& = -g' s \left( 1 + {x^2 c^2(3-2t^2-t^4)\over 8} + \ldots\right)\\
& = - {e M_Z \over M_W}\sqrt{1-{M^2_W \over M^2_Z}}
\left[1+{\cal O}(s^2\,x^4)\right]~.
\label{eq:smii}
\end{align}
Equations (\ref{eq:smi}) and (\ref{eq:smii}), derived from the preceding equations using eqns. (\ref{eq:scdef}), (\ref{eq:mwsq}), (\ref{eq:edef}), and (\ref{eq:zmasssq}) show that the couplings are very
nearly of standard model form; this is a further check of ideal delocalization. 

Since the right-handed light fermion eigenvectors are 
localized entirely at site 2, there are no right-handed couplings of the light fermions to $T_3$ and the right-handed hypercharge coupling of the $Z$ 
is given by
\begin{equation}
g^{Zqq}_{YR} = g' v^2_Z (f^2_R)^2 = g' v^2_Z =  g^{Zqq}_{YL} ~,
\end{equation}
where the last equality comes from eqn. (\ref{eq:smiix}). 

For ideally delocalized light fermions, therefore, we find the $Z$-couplings are given by
the standard-model like expression
\begin{equation}
{e M_Z \over M_W \sqrt{1-{M^2_W \over M^2_Z}}}
\left(T_3 \, P_L -\left(1-{M^2_W \over M^2_Z}\right) Q \right)
\left[1+{\cal O}(s^2\,x^4)\right]~,
\end{equation}
where $P_L$ is the left-handed chirality projection operator.

\subsection{Top and bottom quark couplings to the $Z$ boson}

The left-handed coupling of the top-quark to $T_3$ is
\begin{align}
g^{Ztt}_{3L} &= g (t^0_L)^2 v^0_Z + \tilde{g} (t^1_L)^2 v^1_Z \\
&= g^{Zqq}_{3L} \left(1+{\varepsilon^2_{tR}(2+\varepsilon^2_{tR}) \over 4\,c^2 (1+\varepsilon^2_{tR})^2} x^2+\ldots\right)~.
\end{align}
Note that a similar expression holds for the bottom quark, with $\varepsilon_{tR} \to \varepsilon_{bR}$
and therefore, from eqn. (\ref{eq:epsilonbRlimit}),  the tree-level
corrections to the partial width $\Gamma(Z \to b \bar{b})$ are proportional to
$\varepsilon^2_{bR}/2 < 0.01\%$.
From eqns. (\ref{eq:leftZlag}) and (\ref{eq:rightZlag}), we see that the left- and right-handed top-quark couplings to $Y$ turn out to be the same as those for the other quarks
\begin{align}
g^{Ztt}_{YL} & = g' v^2_Z \left[(t^0_L)^2 + (t^1_L)^2\right] = g^{Zqq}_{YL}\\
g^{Ztt}_{YR} &= g' v^2_Z \left[(t^1_R)^2 + (t^2_R)^2\right] = g^{Zqq}_{YR}~.
\end{align} 
We may also compute the right-handed $T_3$ couplings of the top-quark
\begin{align}
g^{Ztt}_{3R} & = (\tilde{g} v^1_Z-g' v^2_Z)  (t^1_R)^2  \\
& = {g  \over 2c} {\varepsilon^2_{tR}  \over 1+\varepsilon^2_{tR}} (1+{\cal O}(x^2))~.
\end{align}

The $T_3$ couplings of the $Z$ to a pair of heavy-partner fermions or an off-diagonal pair are given in Table \ref{tab:one}.  From the form of eqns. (\ref{eq:leftZlag}) and (\ref{eq:rightZlag}),
we see that the hypercharge couplings of the $Z$ to a pair of left-handed or right-handed heavy-partner fermions follow the pattern of the ordinary fermions:
\begin{equation}
g^{ZQQ}_{YR} = g' v^2_Z =  g^{ZQQ}_{YL}~,
\end{equation}
and the hypercharge coupling of the $Z$ to an off-diagonal (flavor-conserving) $Qq$ pair always vanishes
\begin{equation}
g^{ZQq}_{YL} = g^{ZQq}_{YR} = 0~,
\end{equation}
because the $Q$ and $q$ wavefunctions are orthogonal.

\TABLE[t]
{\begin{tabular}{|c||c|c|}\hline
coupling & calculated as & strength \\ \hline
$g^{Ztt}_{3L}$ & $g_0 v^0_Z (t^0_L)^2 + g_1 v^1_Z (t^1_L)^2$
  & $ cg - \frac{1}{8}c^3 g (3+6t^2-t^4)\, x^2 + {g\varepsilon^2_{tR}(2+\varepsilon^2_{tR}) 
      \over 4\,c (1+\varepsilon^2_{tR})^2}\,  x^2$\\ \hline
$g^{Ztt}_{3R}$ & $(g_1 v^1_Z - g_2 v^2_Z) (t^1_R)^2$
  &$  \dfrac{g \varepsilon_{tR}^2}{2c(\varepsilon_{tR}^2+1)} \left(
      1 +\dfrac{
          -3(\varepsilon_{tR}^2+1)^2
          +8c^2\varepsilon_{tR}^2(\varepsilon_{tR}^2+2)
          -4c^4(\varepsilon_{tR}^2+1)^2
      }{8c^2(\varepsilon_{tR}^2+1)^2} x^2
    \right)  $\\ \hline
$g^{ZTT}_{3L}$ & $g_0 v^0_Z (T^0_L)^2 + g_1 v^1_Z (T^1_L)^2$
   &$ -\frac{1}{2} c g \left(t^2-1\right) + \frac{c g \left(4 \left(t^2+1\right)-c^2
   \left(\varepsilon_{tR}^2+1\right)^2 \left(t^2-1\right)^3\right)}{16
   \left(\varepsilon_{tR}^2+1\right)^2}\, x^2  $\\ \hline
$g^{ZTT}_{3R}$ & $ (g_1 v^1_Z - g_2 v^2_Z) (T^1_R)^2$
  &$ \frac{g}{2c(\varepsilon_{tR}^2+1)}
   + g\frac{\left(-3(\varepsilon_{tR}^2+1)^2
            +8c^2(\varepsilon_{tR}^4 + 3\varepsilon_{tR}^2 +1)
                     - 4c^4(\varepsilon_{tR}^2+1)^2\right)}
       {16c^3(\varepsilon_{tR}^2+1)^3} x^2 $\\ \hline
$g^{ZtT}_{3L}$ & $g_0 v^0_Z t^0_L T^0_L + g_1 v^1_Z  t^1_L T^1_L$
   &$  \frac{g}{2\sqrt{2}c(\varepsilon_{tR}^2+1)} x
   + g\frac{\left((\varepsilon_{tR}^2+1)^2
            +c^2(\varepsilon_{tR}^4 + 6\varepsilon_{tR}^2 -3)
                     - 4c^4(\varepsilon_{tR}^2+1)^2\right)}
       {16\sqrt{2}c^3(\varepsilon_{tR}^2+1)^3} x^3$\\ \hline
$g^{ZtT}_{3R}$& $(g_1 v^1_Z - g_2 v^2_Z) t^1_R T^1_R$
  &$  \frac{g \varepsilon_{tR}}{2c(\varepsilon_{tR}^2+1)}
   + g \varepsilon_{tR} \frac{\left(-3(\varepsilon_{tR}^2+1)^2
            +4c^2(2\varepsilon_{tR}^4 + 5\varepsilon_{tR}^2 +1)
                     - 4c^4(\varepsilon_{tR}^2+1)^2\right)}
       {16c^3(\varepsilon_{tR}^2+1)^3} x^2 $\\ \hline
\end{tabular}
\caption{Strength of the $T_3$ portion of the $Z$ coupling to top flavored fermions in the three-site model to order $x^3$. The $\varepsilon_{tR} \to \varepsilon_{fR}$ limit of a top-flavor coupling is the corresponding coupling of flavor $f$.}
\label{tab:one}}

%%%%%%%%%%%%%%%%%%%%%%%%%%%%%%%%%%%%
%%%%%%%%%%%%%%%%%%%%%%%%%%%%%%%%%%%%
\section{Implications of Multiple Gauge Boson Couplings }

%%%%%%%%%%%%%%%%
%%%%%%%%%%%%%%%%
\subsection{$ZWW$ Vertex and $\epsilon_L$}

Experimental constraints on the $ZWW$ vertex in the three-site model turn out
to provide useful bounds on the fermion delocalization parameter $\epsilon_L$.

To leading order, in the absence of CP-violation, the triple gauge boson vertices
may be written in the Hagiwara-Peccei-Zeppenfeld-Hikasa triple-gauge-vertex 
notation \cite{Hagiwara:1986vm}
\begin{eqnarray}
{\cal L}_{TGV} & = & -ie\frac{c_Z }{s_Z}\left[1+\Delta\kappa_Z\right] W^+_\mu W^-_\nu Z^{\mu\nu}
- ie \left[1+\Delta \kappa_\gamma\right] W^+_\mu W^-_\nu A^{\mu\nu} \cr
&-& i e \frac{c_Z}{s_Z} \left[ 1+\Delta g^Z_1\right](W^{+\mu\nu}W^-_\mu - W^{-\mu\nu} W^+_\mu)Z_\nu 
\label{eq:tgvlag} \\
&-& ie (W^{+\mu\nu}W^-_\mu - W^{-\mu\nu} W^+_\mu) A_\nu~, \nonumber
\end{eqnarray}
where the two-index tensors denote the Lorentz field-strength
tensor of the corresponding field. In the standard model, 
$\Delta\kappa_Z = \Delta\kappa_\gamma = \Delta g^Z_1 \equiv 0$.

As noted in ref. \cite{Chivukula:2005ji}, in any vector-resonance model, such as the Higgsless models considered here, the interactions (\ref{eq:tgvlag}) come from re-expressing the nonabelian couplings in the kinetic
energy terms in the original Lagrangian in terms of the mass-eignestate fields. 
In this case one obtains equal contributions to the deviations of the first and third terms, and the second and fourth terms in eqn. (\ref{eq:tgvlag}). In addition the coefficient of the fourth term is fixed by electromagnetic
gauge-invariance, and therefore in these models we find
\begin{equation}
\Delta \kappa_\gamma \equiv 0 \ \ \ \  \ \Delta\kappa_Z \equiv \Delta g^Z_1~.
\label{eq:tgv_relations}
\end{equation}

Computing the $ZWW$ coupling explicitly in the three-site model \footnote{See also the BESS results \cite{Anichini:1994xx,Casalbuoni:1996pe}.} yields
\begin{align}
g_{ZWW} & = g(v^0_W)^2 v^0_Z + \tilde{g} (v^1_W)^2 v^1_Z \\
& = g\,c\left( 1-{x^2c^2(1+2t^2-t^4) \over 4} + \ldots \right) \\
&= e\frac{c_Z}{s_Z}\left(1 + \frac{1}{8 c^2}x^2 + O(x^4)\right)\label{eq:ninethre}\\
& = g^{Zqq}_{3L} \left(1+{x^2 \over 8 c^2} + \ldots\right)~,
\end{align}
where eqn. (\ref{eq:ninethre}) is derived using (\ref{eq:Delta}). Hence we compute
\begin{equation}
\Delta g^Z_1 = \Delta \kappa_Z = {x^2 \over 8 c^2}>0~.
\label{eq:dgz1}
\end{equation}
The 95\% C.L. upper limit from LEP-II is $\Delta g^Z_1 < 0.028$   \cite{LEPEWWG}.
Approximating $c^2 \approx \cos^2\theta_W \approx 0.77$, we find the bound on $x$
\begin{equation}
x \le 0.42 \sqrt{\Delta g^Z_1 \over 0.028}~,
\end{equation}
and hence, from eqn. (\ref{eq:mwratio}),
\begin{equation}
M_{W'} \approx {2\over x} M_W \ge 380\,{\rm GeV}\, \sqrt{0.028 \over \Delta g^Z_1}~.
\label{eq:mwprimegzww}
\end{equation}
From eqn. (\ref{eq:ideal}), therefore, we can write
\begin{equation}
\varepsilon_L = {m \over M} \approx {x \over \sqrt{2}} \approx
0.30 \left({380\,{\rm GeV} \over M_{W'}}\right)~.
\label{eq:epsilonLconstraint}
\end{equation}

Finally, we recall that, in the absence of a Higgs boson, $W_L W_L$ spin-0 isospin-0 scattering would
violate unitarity at a scale of $\sqrt{8\pi} v$ and that exchange of the heavy electroweak bosons is what unitarizes $WW$ scattering in Higgsless models.  Hence,  $M_{W'} \le 1.2$ TeV in the three-site model.  This constrains $\varepsilon_L$ to lie in the range
\begin{equation}
0.095 \le \varepsilon_L \le 0.30~.
\label{eq:epsilonLrange}
\end{equation}

From eqn. (\ref{eq:mwheavysq}) the bounds on $M_{W'}$ may be
translated directly to bounds on the size of $\tilde{g}$, yielding
\begin{equation}
0.19 < \frac{\tilde{g}^2}{4\pi} < 1.9 \ll 4 \pi~.
\end{equation}
Hence we see that $SU(2)_1$ is moderately strongly coupled, with
radiative corrections (which are proportional to  $\tilde{g}^2/(4\pi)^2$)
of order 20\%.

%%%%%%%%%%%%%%%%
%%%%%%%%%%%%%%%%
\subsection{Comparison to the Continuum Model}

The three-site model may be viewed as an extremely deconstructed version of the  model studied in  ref.~\cite{Chivukula:2005ji}: a five-dimensional flat-space $SU(2)_A \otimes SU(2)_B$ gauge theory with ideally-delocalized fermions. Hence it is interesting to compare the values of the multiple gauge boson and chiral Lagrangian parameters obtained for the two cases.  

The limit of the continuum model that is related to the three-site model has the bulk gauge couplings of  $SU(2)_A$ and $SU(2)_B$, $g_{5A}$ and $g_{5B}$,  equal to one another; in the notation of ref. \cite{Chivukula:2005ji}, $\kappa = \frac{g_{5B}^2}{g_{5A}^2} = 1$.   Then, we may express both the values of the chiral Lagrangian parameters $\alpha_i$ for the three-site model (see Appendix A) and those from ref. \cite{Chivukula:2005ji} in terms of the mass ratio $M_W^2 / M_{W'}^2$, as shown in Table ~\ref{tab:comp2}.  Note that $\Delta g_{WWWW}$ in the table is defined by
\begin{equation}
g_{WWWW} = \frac{e^2}{s_Z^2}
 \left[1 + \Delta g_{WWWW}\right].
\label{eq:Delta_g_WWWW}
\end{equation}
If we fix the value of $\left(\frac{M_W^2}{M_{W'}^2}\right)$, the 
quantities listed in Table~\ref{tab:comp2} for the 
three-site model are about 70\% as large as those for
the continuum model.

\TABLE[t]
{
 \begin{tabular}{|c||c|c|}  \hline
&Three-site model&Continuum model\\ \hline 
$\Delta g_1^Z=\Delta \kappa_Z$& $\frac{1}{2 c^2}\left(\frac{M_W^2}{M_{W'}^2}\right)$&
$\frac{\pi^2}{12 c^2}\left(\frac{M_W^2}{M_{W'}^2}\right)$\\ \hline
$\Delta g_{WWWW}$&$\frac{5}{4}\left(\frac{M_W^2}{M_{W'}^2}\right)$&
$\frac{\pi^2}{5}\left(\frac{M_W^2}{M_{W'}^2}\right)$\\ \hline
$e^2\alpha_1$&$0$&$0$\\ \hline
$e^2\alpha_2 = - e^2 \alpha_3$&$-\frac{s^2}{2}\left(\frac{M_W^2}{M_{W'}^2}\right)$&$-\frac{\pi^2 s^2}{12}\left(\frac{M_W^2}{M_{W'}^2}\right)$\\ \hline
$e^2\alpha_4 = - e^2 \alpha_5$&$ \frac{s^2}{4}\left(\frac{M_W^2}{M_{W'}^2}\right)$&$ \frac{\pi^2 s^2}{30}\left(\frac{M_W^2}{M_{W'}^2}\right)$\\ \hline
 \end{tabular}
\caption{Quantities related to 
multi-gauge-boson vertices and chiral Lagrangian parameters in the three-site model and 
the continuum model of Ref~\cite{Chivukula:2005ji}.}
\label{tab:comp2}
}

%%%%%%%%%%%%%%%%%%%%%%%%%%%%%%%%%%%%
%%%%%%%%%%%%%%%%%%%%%%%%%%%%%%%%%%%%
\section{Phenomenological Bounds}

%%%%%%%%%%%%%%%%
%%%%%%%%%%%%%%%%
\subsection{Corrections to $\alpha T$}

\EPSFIGURE[ht]
{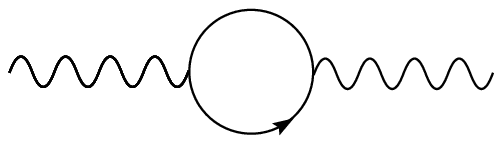,width=0.3\textwidth}
{One-loop contributions to $\Delta \rho$ arise from the differences
in the vacuum polarization diagram for the $W^3$ versus $W^{1,2}$. We compute the leading
contribution in the limit $\varepsilon_L \to 0$ and $m_b \to 0$ (and $g' \to 0$).\label{fig:two}
}

At leading order, there are two isospin-violating 
parameters \cite{Barbieri:2004qk,Chivukula:2004af} of interest. These may
be chosen \cite{Chivukula:2004af} to be $\Delta \rho$, the deviation from one of
the ratio of the strengths of low-energy isotriplet neutral- and
charged-current neutrino interactions, and $\alpha T$, isospin violating
corrections to the masses of the electroweak bosons
\cite{Peskin:1992sw,Altarelli:1990zd,Altarelli:1991fk}. Due to the custodial
symmetry present in the limit $g' \to 0$, at tree-level $\Delta\rho$ is
always equal to zero and $\alpha T$ is small (${\cal O}(x^4)$) 
in models of this kind \cite{Chivukula:2004pk,SekharChivukula:2004mu}.

The existence of the new $T$ and $B$ fermions, the heavy partners of the top and
bottom, gives rise to new one-loop contributions to $\alpha T$, as illustrated in figure \ref{fig:two}. 
In principle, it is the sum of the SM ($t$ and $b$ quark loops) and new physics contributions that is finite.  However, we note that the standard model contribution vanishes in the limit $\varepsilon_L \to 0$ (and $g' \to 0$) since the $t$ and $b$ quark masses are then equal (both vanish, per eqn. (\ref{eq:mtone})).  Since $\varepsilon_L$ respects custodial symmetry, and fermionic custodial symmetry violation is encoded in the $\varepsilon_{fR}$ coefficients, we may obtain the leading contribution to $\alpha T$ from
the new physics by performing the calculation in the $\varepsilon_L \to 0$ limit.  We obtain 
\begin{equation}
\alpha T
\approx {1 \over 16 \pi^2}{{m'_t}^4 \over M^2 v^2} = {1 \over 16 \pi^2} 
{\varepsilon^4_{tR} M^2 \over v^2}~.
\label{eq:deltarho}
\end{equation}

Since $g' \neq 0$, there are also isospin-violating corrections at one-loop in the gauge
sector which yield corrections to $\alpha T$ of order $\alpha/4\pi$. The $M_{W'}$
dependence of the largest of these corrections, which are proportional to 
$\log(M^2_{W'}/M^2_W)$, exactly matches \cite{Matsuzaki:2006wn} the Higgs-boson mass dependence
of isospin-violating contributions at one-loop in the standard model. 
Hence in the three site model, to leading-log 
approximation, the  role of the Higgs boson is largely played by the $W'$.\footnote{At one loop, a heavy SM higgs boson requires an additional positive contribution to $\alpha T$ to bring it into agreement with precision electroweak data.  To the extent that the tree-level values of $\alpha S$ and $\alpha T$ are precisely zero, similar considerations can allow one to set a lower bound on $\alpha T$ in the three-site model as a function of $M_{W'}$ \cite{inprog}.  This, in turn, would provide an upper bound on the Dirac mass of the heavy fermions.}

The phenomenological bounds on the value of $\alpha T$ depend (since
they include the one-loop standard model corrections) 
on the reference Higgs mass chosen. 
We are therefore interested in the bounds on $\alpha T$ corresponding to
Higgs masses between about 380 GeV (from eqn. (\ref{eq:mwprimegzww}))
and the unitarity bound 1.2 TeV. Current bounds (see, for example, Langacker and Erler 
in \cite{Eidelman:2004wy}) yield (approximately) $\alpha T \le 2.5 \times 10^{-3}$,
at 90\% C.L., assuming the existence of a moderately heavy  (340 GeV) Higgs boson, while
it  is relaxed to approximately $\alpha T \le 5 \times 10^{-3}$ 
in the case of a heavy (1000 GeV) Higgs boson. We therefore expect that the
upper bound on  $\alpha T$ in the three site model varies from approximately 
$2.5 \times 10^{-3}$ to  $5 \times 10^{-3}$.
We may rewrite 
eqn. (\ref{eq:deltarho}) as
\begin{equation}
\varepsilon_{tR} = 0.79 \, \left({\alpha T \over 2.5 \times 10^{-3}}\right)^{1/4}
\left({v\over M}\right)^{1/2}~.
\label{eq:epsilonRconstraint}
\end{equation}
In what follows,
we will quote limits on the parameters of the model for both of these values.
For $\alpha T = 5 \times 10^{-3}$, we find the upper bound
\begin{equation}
\varepsilon_{tR} < 0.94 \left({v\over M}\right)^{1/2}~.
\label{eq:simpleepsilonRconstraint}
\end{equation}
As we shall see shortly, this bound is stronger than the one derived from $b\to s\gamma$.

%%%%%%%%%%%%%%%%
%%%%%%%%%%%%%%%%
\subsection{Bounds on $M$}

Our upper limit on $\varepsilon_{tR}$ and our knowledge of the top quark mass allow us to derive a lower bound on $M$.  Our expression (\ref{eq:mt}) for $m_t$ reminds us that 
\begin{equation}
m_t \approx {m m'_t \over \sqrt{M^2 + {m'_t}^2}} =
{\varepsilon_L \varepsilon_{tR} M\over \sqrt{1+\varepsilon^2_{tR}}}~.
\label{eq:mtequation}
\end{equation}
For a given value of $M$, the existence of an upper bound on $\varepsilon_{tR}$ implies that there
is a  {\it smallest} allowed value of $\varepsilon_L$, which we denote $\varepsilon^*_L$ %
\begin{equation}
\varepsilon^*_L =  1.26\left({2.5 \times 10^{-3}\over \alpha T }\right)^{1/4}
{m_t \over \sqrt{vM}}\,\sqrt{1+0.63\left({\alpha T \over 2.5 \times 10^{-3}}\right)^{1/2}{v\over M}}~.
\label{eq:smallepsilonL}
\end{equation}
Since eqn. (\ref{eq:epsilonLrange}) requires $\varepsilon^*_L < 0.30$, for 
$\alpha T=2.5 \times 10^{-3}$ we find that  $M$ must be greater than 2.3 TeV,
and for $\alpha T = 5 \times 10^{-3}$ we find that $M$ must be greater than
1.8 TeV.

%\begin{figure}[ht]
%\centering
%\includegraphics[width=0.45\textwidth]{graphics_new.eps}
%\includegraphics[width=0.45\textwidth]{graph2_new.eps}
%\label{fig:three}
%\caption{ (left) and
%the corresponding values of $M$ and $M_{W'}$ (right).
% in the left figure, and .
%}
%\end{figure}

\DOUBLEFIGURE[ht]{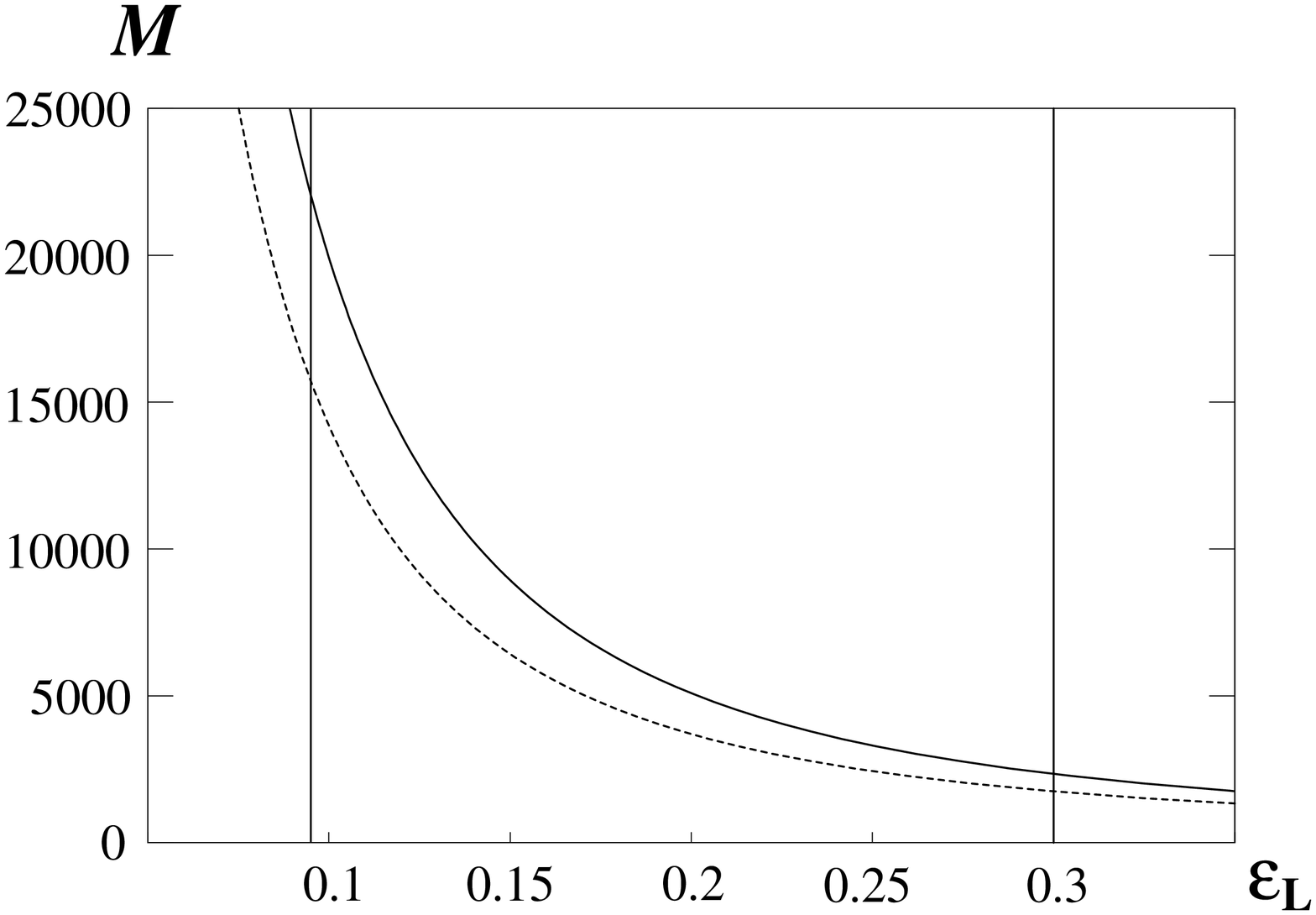,width=0.45\textwidth }{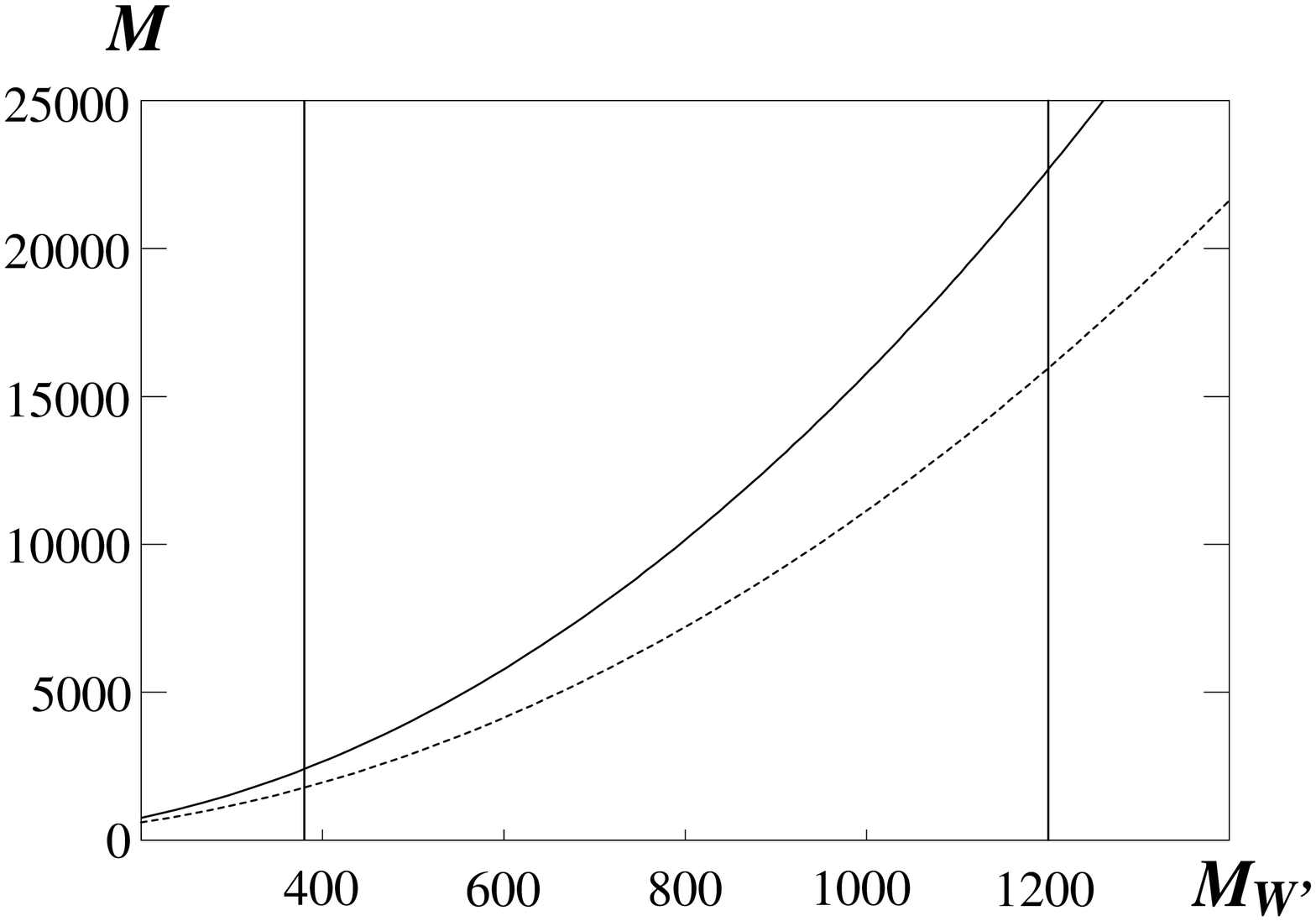,width=0.45\textwidth }{Phenomenologically acceptable values of $M$ in GeV and $\varepsilon_L$ for $\alpha T=2.5 \times 10^{-3}$ (solid curve) and $5 \times 10^{-3}$ (dashed curve). The region bounded by the lines $0.095 < \varepsilon_L < 0.30$, and above the appropriate curve is allowed. For a given $M$ and
$\varepsilon_L$, 
the value of $\varepsilon_{tR}$ is determined by eqn. (\protect\ref{eq:mtequation}).
As discussed in the text, naive dimensional analysis implies $M < 46$ TeV.}
{Phenomenologically acceptable values of $M$ and $M_{W'}$ in GeV for $\alpha T=2.5 \times 10^{-3}$ (solid curve) and $5 \times 10^{-3}$ (dashed curve). The region bounded by the lines  $380\, {\rm GeV}< M_{W'} < 1200\,{\rm GeV}$ and above the appropriate curve is allowed. 
For a given $M$ and $M_{W'}$ (see eqn. (\protect\ref{eq:epsilonLconstraint})), 
the value of $\varepsilon_{tR}$ is determined by eqn. (\protect\ref{eq:mtequation}).
As discussed in the text, naive dimensional analysis implies $M < 46$ TeV.}

Several additional consequences follow.  
Using $M>1.8$ TeV  and the bound in eqn. (\ref{eq:simpleepsilonRconstraint}), we see that 
$\varepsilon_{tR} < 0.35$, which supersedes the  
$b \to s \gamma$ constraint, eqn. (\ref{eq:bsgammaconstraint}), as promised above.
For $\alpha T = 2.5 \times 10^{-3}$, as $M$ grows above its minimum value of $2.3$ TeV, according to eqn. (\ref{eq:smallepsilonL}) 
the value of
$\varepsilon^*_L$ will falls -- reaching  the lower bound of 0.095 (eqn. (\ref{eq:epsilonLrange})) when
$M\approx 22$ TeV. For values of $M$ greater than 22 TeV (and fixed $m_t$), the entire range of
$0.095 < \varepsilon_L < 0.30$ remains accessible if  $\varepsilon_{tR}$ is smaller than its maximum value (which, for $\alpha T=2.5 \times 10^{-3}$, is 0.26). The joint range of allowed $\varepsilon_L$ and $M$ for both values of $\alpha T$ is
summarized in figures 3 and 4.

In the simplest continuum models in which the 5th dimension is a flat interval, 
the mass of the first KK fermion resonance is approximately half of the mass of the 
first gauge-boson KK resonance.  Due to the chiral boundary
conditions on the fermions, Dirichlet at one boundary and Neumann at the other,
the lowest KK fermion mode has a wavelength of twice the size of the 5-d interval. Phenomenologically,
this situation is disfavored --- and it has been suggested that this may be addressed by having the
fermions ``feel" a smaller size for the 5-d interval than the gauge-bosons 
\cite{Cacciapaglia:2005pa,Foadi:2005hz}. Following \cite{Foadi:2005hz}, the parameter 
which measures this enhancement is then given by
\begin{equation}
2\,{m_{t*} \over M_{W'}}~,
\end{equation}
and, for the three-site model, we find its minimum value  is
about 12 for $\alpha T = 2.5 \times 10^{-3}$ and 7 for $\alpha T= 5 \times 10^{-3}$. In other words, viewing the three site model as the
deconstruction of a continuum one, the bulk fermion fields behave as though the 
5-d interval is at least twelve or seven times smaller than do the gauge-bosons.
%In other words, if even if the $W'$ is as light as $ 380$ GeV, the heavy fermions will
%have masses of at least 4.6 TeV.

It is interesting to ask what {\it upper} bound exists on $M$.  From the expression for
the fermion mass matrix in eqns. (\ref{eq:fermmat})  we have
\begin{equation}
 M = {\sqrt{2} \lambda v \over \varepsilon_L}.
 \label{eq:Mbound}
\end{equation}
Eqn. (\ref{eq:epsilonLrange}) requires $\varepsilon_L \geq 0.095$, and from naive dimensional
analysis or, equivalently, perturbative unitarity, we expect $\lambda \le 4 \pi$. Hence,  $M < 46 $ TeV. A more sophisticated analysis could be done by imposing
unitarity of $WW \to t \bar{t}$, as in \cite{Foadi:2005hz} (see also, footnote 7).

%%%%%%%%%%%%%%%%%%%%%%%%%%%%%%%%%%%%
%%%%%%%%%%%%%%%%%%%%%%%%%%%%%%%%%%%%

\section{Decoupling and $Z\to b\bar{b}$ with a Toy UV Completion}

\label{decouplesection}

In the analysis above we have argued that the ``bulk fermion" Dirac mass $M$ in
the three site model must be large, between 1.8 and 46  TeV. Such a mass is 
potentially much larger than   $4\pi \sqrt{2}\, v \approx 4.3$ TeV, the largest mass which can arise 
from the symmetry breaking encoded by the link fields. By contrast, the nonlinear sigma
model link fields have so far been described by an effective chiral lagrangian which
is valid only at energies {\it less} than of order $4\pi \sqrt{2}\, v$. In order to discuss
the three-site model in the large-$M$ limit, therefore, one must consider the question
in the context of a theory which is consistent to much higher scales -- {\it e.g.}
a renormalizable one. The situation here is similar to the consistent analysis
\cite{Golden:1994pj} of the Appelquist-Chanowitz bound \cite{Appelquist:1987cf}.

The simplest possible renormalizable extention of the three site
model is formed by promoting the link fields in
figure \ref{fig:one} to linear sigma-model fields.  Here one introduces
two additional singlet fields $H_i$ ($i=1,2$) and considers the
matrix fields
\begin{equation}
\Phi_{i} = {(H_{i} +f_{i})\over 2}\,  \Sigma_{i}~,
\end{equation}
which transform as $(2,\bar{2})$ under the appropriate
$SU(2)$'s, and which have the kinetic energy terms
\begin{equation}
{\rm Tr} \left(D^\mu \Phi^\dagger_{i} D_\mu \Phi_{i}\right)
\to {1\over 2} \partial^\mu H_i \partial_\mu H_i  + {f^2_i \over 4} {\rm Tr}
\left(D^\mu \Sigma^\dagger_i D_\mu \Sigma_i\right)~.
\end{equation}
The most general renormalizable potential for the fields $\Phi_{1,2}$ will result
in mixing between the fields $H_{1,2}$, which will therefore not be mass eigenstates.
For the purposes of this note, however, this will not be relevant --- we will require that,
consistent with dimensional analysis, the masses of the ``Higgs" are bounded
by $4\pi \sqrt{2} v$.

For completeness, in this section we will carry the explicit 
dependence on $f_{1,2}$, although in practice we always have
in mind $f_1 \simeq f_2 \simeq \sqrt{2} v$ (as in eqn. (\ref{eq:equalfs})).
We continue to work in the limit in which 
$x={g_0/g_1} \ll 1$ and $y={g_2/g_1}\ll 1$.

In this linear sigma model version of the three-site model, the 
Yukawa couplings and fermion mass term are of the form below, which is
the natural extension of eqn. (\ref{eq:yukawa})
\begin{align}
{\cal L}_f = &  \varepsilon_L M \left(1+{H_1\over f_1}\right)  \bar{\psi}_{L0} \Sigma_1 \psi_{R1}
 + M \bar{\psi}_{R1} \psi_{L1}  + \cr
 &  M \left(1+{H_2\over f_2}\right) \bar{\psi}_{L1} \Sigma_2
\begin{pmatrix}
\varepsilon_{uR} & \\
& \varepsilon_{dR}
\end{pmatrix}
\begin{pmatrix}
u_{R2} \\
d_{R2}
\end{pmatrix}
+ h.c.
\label{eq:yukawaM}
\end{align}
Although the Yukawa couplings are written in terms of the Dirac
mass $M$ for convenience, we do impose (see the discussion surrounding
eqn. (\ref{eq:Mbound})) the consistency constraint
\begin{equation}
{\varepsilon_L M \over f_1}~,\  {(\varepsilon_{uR},\,\varepsilon_{dR})\, M\over f_2} < 4\pi~,
\label{eq:consistencyM}
\end{equation}
on the size of the allowed Yukawa couplings.

\subsection{The Large $M$ Effective Theory}

We now consider the large-$M$ limit. Due to the decoupling theorem \cite{Appelquist:1974tg},
the effects of the bulk (i.e. site-1) fermion on low-energy parameters must be suppressed
by powers of $M$.  Due to the parameterization of the couplings chosen
in eqn. (\ref{eq:yukawaM}), the form of the operators in the low-energy effective theory
may not obviously appear to be suppressed by $M$ when written in terms of the parameters
$\varepsilon_L$ and $\varepsilon_{uR,dR}$. Nonetheless, 
because of the constraints of eqn. (\ref{eq:consistencyM}), the effects of the bulk
fermion always formally decouple in the $M \to \infty$ limit \cite{Appelquist:1974tg}.  We will now look at light fermion masses, the coupling of delocalized light fermions to gauge bosons, and $\alpha T$ in the large-$M$ limit and see how the results compare with our previous findings.

\subsubsection{Light Fermion Masses}

\begin{figure}[ht]
\centering
\begin{center}
%\fcolorbox{white}{white}
{
  \begin{picture}(210,104) (17,-23)
    \SetWidth{0.5}
     \SetColor{Black}
     \ArrowLine(61,-17)(91,13)
     \ArrowLine(91,13)(150,13)
     \ArrowLine(150,13)(181,-17)
     \DashArrowLine(61,44)(91,14){2}
     \DashArrowLine(182,44)(151,15){2}
     \Text(191,-23)[lb]{\Large{\Black{$\psi_{L0}$}}}
     \Text(0,-23)[lb]{\Large{\Black{$u_{R2},\, d_{R2}$}}}
     \Text(197,53)[lb]{\Large{\Black{$\Phi_1$}}}
     \Text(32,53)[lb]{\Large{\Black{$\Phi_2$}}}
   \end{picture}
 }
\end{center}
\label{fig:twoM}
\caption{Mass-mixing diagram which yields the operator in eqn. (\protect\ref{eq:seesawmassM})
when integrating out the bulk fermion (interior fermion line)  at tree-level.}
\end{figure}
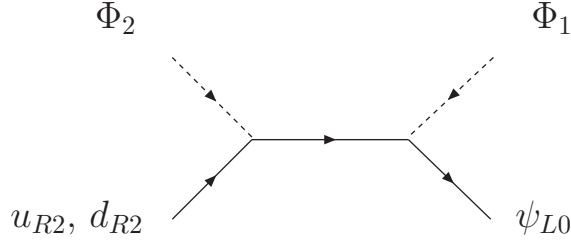

The masses of the ordinary fermions arise in the large-$M$ limit when we consider the diagram connecting
left-handed (site-0) and right-handed (site-2) brane fermions in fig. 5 and then integrate out the intervening site-1 bulk fermion at tree level.  Specifically, this gives rise to the operator:
\begin{align}
{\cal L}_f' \propto &
{\varepsilon_L M \over f_1 f_2}
\,\bar{\psi}_{L0} \Phi _1 \Phi_2
\begin{pmatrix}
\varepsilon_{uR} & \\
& \varepsilon_{dR}
\end{pmatrix}
\begin{pmatrix}
u_{R2} \\
d_{R2}
\end{pmatrix}
+ h.c~.
\label{eq:seesawmassM}
\\
\propto & 
\,\varepsilon_L\,M \left(1+{H_1\over f_1}\right)
\left(1+{H_2\over f_2}\right)
\,\bar{\psi}_{L0} \Sigma _1 \Sigma_2
\begin{pmatrix}
\varepsilon_{uR} & \\
& \varepsilon_{dR}
\end{pmatrix}
\begin{pmatrix}
u_{R2} \\
d_{R2}
\end{pmatrix}
+ h.c~.
\end{align}
In unitary gauge (with $\Sigma_i = I$) the leading term provides the up-type fermion with a mass of the typical seesaw form $m_u \propto \varepsilon_L \varepsilon_{uR} M$ that agrees with eqn. (\ref{eq:mtone}), 
and similarly for the down-type fermion.   The overall power of $M$ results from two powers of $M$ in the Yukawa couplings, and one factor of $1/M$ from the propagator in fig. 5. 

\subsubsection{Ideally Delocalized Fermion Couplings}

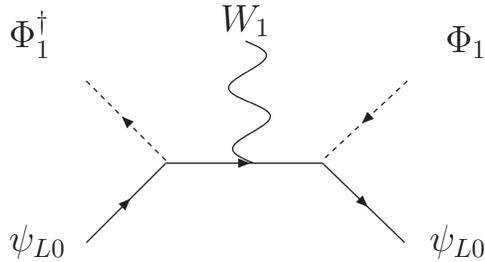
\begin{figure}[ht]
\centering
\begin{center}
 %\fcolorbox{white}{white}
 {
   \begin{picture}(210,115) (17,-12)
     \SetWidth{0.5}
     \SetColor{Black}
     \ArrowLine(61,-6)(91,24)
     \ArrowLine(91,24)(150,24)
     \ArrowLine(150,24)(181,-6)
     \DashArrowLine(91,25)(61,55){2}
     \DashArrowLine(182,55)(151,26){2}
     \Text(191,-12)[lb]{\Large{\Black{$\psi_{L0}$}}}
     \Text(32,-12)[lb]{\Large{\Black{$\psi_{L0}$}}}
     \Text(197,64)[lb]{\Large{\Black{$\Phi_1$}}}
     \Text(32,64)[lb]{\Large{\Black{$\Phi^\dagger_1$}}}
     \Photon(121,70)(124,24){7.5}{2}
     \Text(112,72)[lb]{\Large{\Black{$W_1$}}}
   \end{picture}
 }
\end{center}
\label{fig:threeM}
\caption{Coupling diagram which yields the operator in eqn. (\protect\ref{eq:couplingmixM})
when integrating out the bulk fermion at tree-level.}
\end{figure}

In this limit, it should also be possible to obtain an effective coupling of $\psi_{L0}$ to the $SU(2)$ group at site 1, consistent with light-fermion delocalization.  Indeed, considering fig. 6 and integrating out the bulk  fermion at tree-level induces the operator
\begin{align}
{\cal L}_{Wff}'  \propto  &
{\varepsilon_L^2  \over f^2_1}\, \bar{\psi}_{L0}
\Phi_1 i \slashiii{D} \Phi^\dagger_1 \psi_{L0} ~,
\label{eq:couplingmixM}
\\
\supset &\,
\varepsilon_L^2\, \bar{\psi}_{L0}
\Sigma_1 i \slashiii{D} \Sigma^\dagger_1 \psi_{L0} ~,
\end{align}
which includes (easily visible in unitary gauge) just such an effective coupling.  In this case, 
the two powers of $M$ in the Yukawa couplings are cancelled by
$1/M^2$ from the fermion propagators in fig. 6.  If  we adjust the value of the coefficient $\epsilon_L$ to make the coupling of the light fermions to the $W'$ vanish, we achieve ideal delocalization of the light fermions.  The coupling of the  brane fermions to the bulk gauge
group is then precisely of the form discussed in \cite{Anichini:1994xx,SekharChivukula:2005xm}.

\subsubsection{Deviations in $\alpha T$}

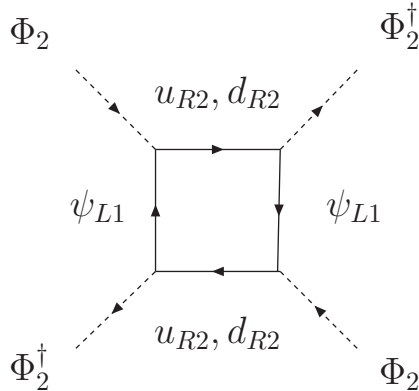
\begin{figure}[ht]
\centering
\begin{center}
 %\fcolorbox{white}{white}
 {
   \begin{picture}(212,166) (58,-75)
     \SetWidth{0.5}
     \SetColor{Black}
     \ArrowLine(135,-30)(135,16)
     \ArrowLine(135,16)(182,16)
     \ArrowLine(182,16)(181,-30)
     \ArrowLine(182,-30)(135,-30)
     \DashArrowLine(182,16)(211,46){2}
     \DashArrowLine(135,-30)(105,-59){2}
     \DashArrowLine(105,46)(135,16){2}
     \DashArrowLine(211,-59)(182,-30){2}
     \Text(220,54)[lb]{\Large{\Black{$\Phi^\dagger_2$}}}
     \Text(80,-75)[lb]{\Large{\Black{$\Phi^\dagger_2$}}}
     \Text(80,54)[lb]{\Large{\Black{$\Phi_2$}}}
     \Text(220,-75)[lb]{\Large{\Black{$\Phi_2$}}}
     \Text(135,31)[lb]{\Large{\Black{$u_{R2},d_{R2}$}}}
     \Text(135,-60)[lb]{\Large{\Black{$u_{R2},d_{R2}$}}}
     \Text(200,-12)[lb]{\Large{\Black{$\psi_{L1}$}}}
     \Text(103,-12)[lb]{\Large{\Black{$\psi_{L1}$}}}
   \end{picture}
 }
\end{center}
\label{fig:fourM}
\caption{Loop Diagram giving the leading contribution to $\alpha T$, as
encoded in the operator of eqn. (\protect\ref{eq:deltarhopM}).}
\end{figure}

We may also check that the size of $\alpha T$ in the large-$M$ limit is consistent with our previous calculation.  As discussed in section 8.1, we expect that the leading contributions from beyond-the-standard-model physics will persist in the $\varepsilon_L \to 0$ limit and will arise, in fact, from the weak isospin violation encoded  in the $\varepsilon_{fR}$. Then the appropriate diagram\footnote{See
\cite{Collins:1999rz} for a similar analysis in the case of the top-quark seesaw model.}  involves only the $ \bar{\psi}_{L1} \Phi_2 (u_{R2}, d_{R2})$ Yukawa couplings in eqn. (\ref{eq:yukawaM}) as shown in fig. 7, and gives rise to the operator\footnote{One may also deduce that this is the leading operator by recalling that  $\alpha T$ violates weak isospin by two units.  An iso-triplet operator would not suffice.}
\begin{align}
{\cal L}'_{\Delta \rho} \propto &
 {M^2\over 16\pi^2f^4_2} \left({\rm Tr }
\begin{pmatrix}
\varepsilon^2_{uR} & \\
& \varepsilon^2_{dR}
\end{pmatrix}
\Phi_2^\dagger D^\mu \Phi_2
\right)^2~,
\label{eq:deltarhopM}
\\
 \propto &
 {M^2\over 16\pi^2} \left({\rm Tr }
\begin{pmatrix}
\varepsilon^2_{uR} & \\
& \varepsilon^2_{dR}
\end{pmatrix}
\Sigma_2^\dagger D^\mu \Sigma_2
\right)^2~.
\end{align}
In unitary gauge, this may be seen to affect the mass of the $Z$ and not that of the $W$. It encodes the very corrections to $\alpha T$ discussed in section 8.1 and eqn. (\ref{eq:deltarho}).
Here the $M^2$ 
arises from four powers of $\varepsilon_R M$ from the couplings and
an overall $1/M^2$ from the convergent loop integral in the diagram.

\subsection{$Z \to b\bar{b}$}

With this background, we may now discuss flavor-dependent corrections
to the process $Z \to b \bar{b}$. We will do so in the limit that $m_b =0$,
and hence $\varepsilon_{bR}=0$. In the large-$M$ limit, therefore,
we are interested in flavor-dependent corrections to the coupling
of the lower-component of $\psi_{L0}$ to the $SU(2)_0$ gauge-bosons.
Furthermore, as we are interested in flavor-nonuniversal contributions, we
are only interested in couplings proportional to $\varepsilon_{tR}$ --- any
contributions depending only on $\varepsilon_L$ will be flavor-universal.

\begin{figure}[ht]
\centering
\begin{center}
 %\fcolorbox{white}{white}
 {
   \begin{picture}(263,135) (58,-94)
     \SetWidth{0.5}
     \SetColor{Black}
     \ArrowLine(90,-91)(136,-45)
     \ArrowLine(136,-45)(224,-45)
     \ArrowLine(224,-45)(271,-91)
     \DashArrowArc(178.62,29.82)(87.85,-119.02,-59.66){10}
     \Photon(151,2)(155,-46){7.5}{2}
     \DashArrowLine(136,-45)(91,0){2}
     \DashArrowLine(180,-45)(180,0){2}
     \DashArrowLine(272,0)(224,-45){2}
     \DashArrowLine(211,1)(211,-46){2}
     \Text(279,-94)[lb]{\Large{\Black{$\psi_{L0}$}}}
     \Text(60,-94)[lb]{\Large{\Black{$\psi_{L0}$}}}
     \Text(178,-73)[lb]{\Large{\Black{$\pi$}}}
     \Text(60,5)[lb]{\Large{\Black{$\Phi^\dagger_1$}}}
     \Text(145,5)[lb]{\Large{\Black{$W_1$}}}
     \Text(175,5)[lb]{\Large{\Black{$\Phi^\dagger_2$}}}
     \Text(206,7)[lb]{\Large{\Black{$\Phi_2$}}}
     \Text(279,7)[lb]{\Large{\Black{$\Phi_1$}}}
   \end{picture}
 }
\end{center}
\caption{Loop Diagram giving leading contribution to the nonuniversal correction
to $Z \to b\bar{b}$. Here $\pi$ corresponds to a quantum charged ``eaten" Goldstone
boson, and the vertex involving the fermions, $\Phi^\dagger_1$ and $\pi$
is to be interpreted using the ``background field" method to preserve chiral invariance.
This diagram, and those like it, give rise to the operator in eqn. (\protect\ref{eq:zbbopM})
in the low-energy theory.}
\label{fig:fiveM}
\end{figure}
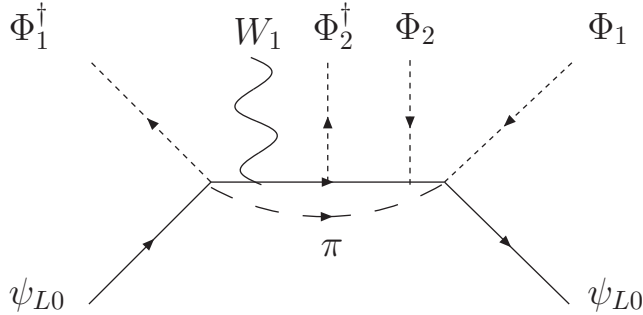

There are no relevant contributions at tree-level, as the neutral gauge-boson
couplings involving $\varepsilon_{tR}$ at tree-level couple to the
upper component of $\psi_{L0}$ -- {\it i.e.} to the top-quark. The leading
contributions arise from diagrams of the form shown in fig. \ref{fig:fiveM}. Note that the
diagram involves the exchange of a charged Goldstone-boson (necessary to
couple to the lower component of $\psi_{L0}$), two couplings proportional
to $\varepsilon_L$ and two proportional to $\varepsilon_{tR}$. This diagram,
and those like it, give rise to the low-energy operator
\begin{equation}
{\cal L}_{Zbb}\propto {\varepsilon^2_L \over 16 \pi^2 f^2_1\, f^2_2}
\sum_a\,
\bar{\psi}_{L0}\left[
 \left({\tau^a \over 2}\right)
\Phi^\dagger_1 \slashiii{D} \Phi^\dagger_2
\begin{pmatrix}
\varepsilon^2_{uR} & \\
& \varepsilon^2_{dR}
\end{pmatrix}
\Phi_2 \Phi_1
\left({\tau^a \over 2}\right)
\right]
\psi_{L0}~.
\label{eq:zbbopM}
\end{equation}
Here four powers of $M$ from the Yukawa couplings are cancelled by
$1/M^4$ from dimensional analysis. 

An operator of this sort gives rise
to a shift in the $Zbb$ coupling of order
\begin{equation}
{\delta g_{Zbb} \over g^{SM}_{Zbb}} \propto 
{\varepsilon^2_L \varepsilon^2_{tR} \over 16 \pi^2}
= {m^2_t \over 16 \pi^2 M^2}~.
\end{equation}
By contrast, the one-loop SM contribution to the $Zb\bar{b}$ coupling \cite{Barbieri:1992nz,Barbieri:1992dq}
is of order $m^2_t / 16 \pi^2 v^2$.
We therefore see that the new corrections to the process $Z\to
b\bar{b}$ arising in the three site model are  likely, even for the lowest possible $M$, to be
negligibly small!$^6$

\begin{figure}[ht]
\centering
\begin{center}
 %\fcolorbox{white}{white}
 {
   \begin{picture}(211,86) (14,-155)
     \SetWidth{0.5}
     \SetColor{Black}
     \Photon(45,-125)(98,-123){7.5}{3}
     \ArrowArc(121,-123)(23.32,149,509)
     \ArrowLine(144,-125)(179,-95)
     \ArrowLine(182,-155)(143,-125)
     \Text(120,-90)[lb]{\Large{\Black{$t$}}}
     \Text(188,-160)[lb]{\Large{\Black{$b$}}}
     \Text(188,-95)[lb]{\Large{\Black{$b$}}}
     \Text(14,-130)[lb]{\Large{\Black{$Z$}}}
   \end{picture}
 }
\end{center}
\caption{A potentially large correction to $Z\to b\bar{b}$ in extra-dimensional
models \cite{Oliver:2002up}. Due to ideal delocalization, however, the $W'tb$ coupling vanishes,
and this contribution is small in the three site model described here.}
\label{eq:fignine}
\end{figure}
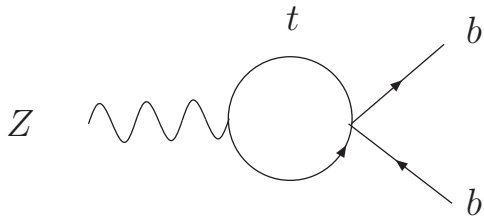

In models with an extra dimension, one might generally be worried about effects which
arise from integrating out the KK modes \cite{Oliver:2002up}, as shown
in fig. \ref{eq:fignine}. Integrating out the heavy $W'$ would lead one to anticipate a relatively large contribution of the form
\begin{equation}
{\delta g_{Zbb} \over g^{SM}_{Zbb}} \simeq {g^2 v^2 \over 16 \pi^2 M^2_{W'}}
\log\left({M^2_{W'} \over m^2_t}\right)~.
\end{equation}
In a theory with ideal delocalization, however, the $W'tb$ coupling vanishes, and
therefore there is no such effect in the three site model described here.

We should also note that the estimate given above provides
only a lower bound on the size of the corrections to the $Zb\bar{b}$
coupling. It is possible that in a truly dynamical model, the 
``extended technicolor"-like physics responsible for generating the Yukawa couplings
can give rise to new contributions \cite{Chivukula:1992ap}.

Finally, it is worth mentioning that 
the situation could be somewhat different in a model with ``Georgi fermions" \cite{Georgi:2005dm}.
In this case, all of the fermion masses arise from dimension-four Yukawa
couplings so that it is not possible to take the large-$M$ limit, even
in principle. Nonetheless, the analysis given here, taking the limit $M \to {\cal O}(4\pi v)$,
shows that the effects on $Z\to b\bar{b}$ are still likely to be quite small.

\section{Conclusions}

The three-site model is a useful tool for illustrating many issues of current interest in Higgsless models: ideal fermion delocalization, precision electroweak corrections, fermion mass generation,  and phenomenological constraints.   Because the Moose describing the model has only one interior $SU(2)$ group, there is, accordingly, only a single triplet of $W'$ and $Z'$ states instead of the infinite tower of triplets present in the continuum limit.  Likewise, there need only be single heavy fermion partner for each of the standard model fermions, instead of a tower of such states.  Because the $W'$ and $Z'$ states are fermiophobic when the light fermions are ideally delocalized, discovering these heavy gauge bosons at a high-energy collider will require careful study of gauge-boson-fusion processes \cite{Chivukula:2005ji,Birkedal:2004au}.  Fortunately, the sparse spectrum and limited number of model parameters should allow this model to be encoded in a Matrix Element Generator program in concert with a Monte Carlo Event Generator for detailed phenomenological investigations.

In this paper, we have discussed the forms of the gauge boson and fermion wavefunctions and their couplings to one another, and then explored the phenomenological implications.  We established the form of the fermion Yukawa couplings required to produce the ideal fermion delocalization that causes tree-level precision electroweak corrections to vanish by making the $W'$ and $Z'$ fermiophobic.  We 
discussed the implications of corrections to multi-gauge-boson vertices for ideal delocalization, and compared the sizes of electroweak chiral Lagrangian parameters in the three-site model with those for the continuum limit.   In addition, we have studied a variety of phenomenological constraints arising from  anomalous gauge couplings,  and from one-loop corrections to $b\to s \gamma$ and the weak-isospin violating parameter $\alpha T$. We found that the extra fermiophobic vector boson states (the analogs of the gauge-boson KK modes in a continuum model) can be reasonably light, with a mass as low as 380 GeV, while the extra (approximately vectorial) quark and lepton states must satisfy 1.8 TeV $\leq M \leq $ 46 TeV.

Because the bulk fermion's Dirac mass $M$ does not arise from electroweak symmetry breaking, its effects on low-energy parameters must decouple.  To investigate this explicitly, we have constructed a 
large-$M$ effective field theory.  Since $M$ lies above the range of validity of the non-renormalizable non-linear sigma model for the link fields,  our analysis employs the
simplest possible UV completion, in which the link fields are given by renormalizable
linear sigma models.  This allows us to construct an effective low-energy theory produced when the bulk femions of mass $M$ are integrated out.
We confirmed that the results in the large-$M$ effective theory for the top-quark 
mass, the gauge-boson couplings required by ideal delocalization, and the
one-loop contribution to $\alpha T$ are precisely those we computed
directly.  We then used the large-$M$ effective theory to estimate the size of the nonuniversal
corrections to the $Zb\bar{b}$ coupling -- and found that these corrections can
be very small, proportional to $m^2_t / 16 \pi^2 M^2$. 

\vspace{1cm}

\centerline{\bf Acknowledgments}

R.S.C., E.H.S., B.C., and  S.D. are supported in part by the US National Science Foundation under
grant  PHY-0354226.  M.T.'s work is supported in part by the JSPS Grant-in-Aid for Scientific Research No.16540226. H.J.H. is supported by Tsinghua University.  M.K. is supported in part by the US National Science Foundation under grant PHY-0354776. R.S.C. and E.H.S. thank the Aspen Center for
Physics for its hospitality while this work was being completed.

\appendix

\section{Four-Point Gauge Vertices and Chiral Lagrangian Parameters}

This appendix gives expressions for Chiral Lagrangian parameters and
the quartic W boson coupling for the three-site model in the notation used in
this paper.   These quantities have previously been derived for the equivalent gauge
sector of the BESS model in \cite{Anichini:1994xx,Casalbuoni:1997bv}.

Of the complete set of 12 CP-conserving operators in the electroweak chiral Lagrangian written down by Longhitano \cite{Appelquist:1980vg,Longhitano:1980iz,Longhitano:1980tm,Appelquist:1980ix,Holdom:1990xq,Falk:1991cm} and Appelquist and Wu \cite{Appelquist:1993ka}, only five apply to Higgsless models such as the three-site model (see ref. \cite{Chivukula:2005ji} for a discussion):
\begin{eqnarray}
{\cal L}_1 &\equiv& \frac12 \alpha_1 g_W g_Y B_{\mu\nu} Tr(T W^{\mu\nu})\\
{\cal L}_2 &\equiv& \frac12 i \alpha_2 g_Y B_{\mu\nu} Tr(T [V^\mu, V^\nu])\\
{\cal L}_3 &\equiv& i \alpha_3 g_W Tr(W_{\mu\nu} [V^\mu, V^\nu])\\
{\cal L}_4 &\equiv& \alpha_4 [ Tr(V^\mu V^\nu)]^2\\
{\cal L}_5 &\equiv& \alpha_5 [Tr(V_\mu V^\mu)]^2~.
\end{eqnarray}
Here $W_{\mu\nu}$, $B_{\mu\nu}$, $T \equiv U\tau_3U^\dagger$ and $V_\mu \equiv (D_\mu U)U^\dagger$ are the basis of the expansion, with
$U$ being the nonlinear sigma-model field\footnote{$SU(2)_W \equiv SU(2)_L$ and $U(1)_Y$ is
identified with the $T_3$ part of $SU(2)_R$.}  arising from $SU(2)_L \otimes SU(2)_R \to
SU(2)_V$. 
An alternative parametrization by Gasser and Leutwyler \cite{Gasser:1983yg} names these coefficients as $\alpha_1 = L_{10}~,\ \alpha_2 = - \frac12 L_{9R}~,\  \alpha_3 = - \frac12 L_{9L}~,\ 
\alpha_4 = L_2~, \  \alpha_5 = L_1$.

The chiral Lagrangian coefficients are related\footnote{$\Delta \kappa_\gamma (=0) = \frac{1}{s^2}(-e^2\alpha_1 + e^2\alpha_2 + e^2\alpha_3)$ is automatically satisfied 
when $\Delta g_1^Z = \Delta \kappa_Z$.}
 to $\alpha S$, the Hagiwara-Peccei-Zeppenfeld-Hikasa \cite{Hagiwara:1986vm} triple-gauge-vertex parameters and the quartic $W$ boson vertex as follows \cite{Chivukula:2005ji}:
\begin{eqnarray}
 \alpha S &=& -(16\pi\alpha)\ \alpha_1, \label{appx-eqa} \\
 \Delta g_1^Z &=& \frac{1}{c^2(c^2-s^2)}e^2\alpha_1
 + \frac{1}{s^2c^2}e^2\alpha_3, \\
 \Delta \kappa_Z &=& \frac{2}{(c^2-s^2)}e^2\alpha_1
 - \frac{1}{c^2}e^2\alpha_2
 + \frac{1}{s^2}e^2\alpha_3, \\
g_{WWWW} &=& \frac{e^2}{s_Z^2}
 \left[1+\frac{2}{(c^2-s^2)}e^2\alpha_1
 + \frac{2}{s^2}e^2\alpha_3
 + \frac{1}{s^2}e^2\alpha_4\right].
 \label{appx-eqd}
\end{eqnarray}

 An expression for $g_{WWWW}$ may be calculated as follows:
\begin{eqnarray}
 g_{WWWW} &=& g^2 (v_W^0)^4 + {\tilde g}^2 (v_W^1)^4,\nonumber \\
 &=& g^2 \left(1-\frac{7}{16}x^2 + O(x^4)\right).
 \end{eqnarray}
 Using eqn. (\ref{eq:Delta}) we may re-express this as
 \begin{equation}
 g_{WWWW} =  \frac{e^2}{s_Z^2}\left(1 + \frac{5}{16}x^2 + O(x^4)\right).
\end{equation}

 Solving the relations in (\ref{appx-eqa}) - (\ref{appx-eqd}) using $S=O(x^4)$, the values of 
$\Delta g_1^Z$ and $\Delta \kappa_Z$ from the main text, and $g_{WWWW}$ as above, we obtain:
\begin{equation}
 e^2\alpha_1 = O(x^4),
\end{equation}
\begin{equation}
 e^2\alpha_2 = - e^2\alpha_3 = -\frac{s^2}{8}x^2 + O(x^4),
\end{equation}
\begin{equation}
 e^2\alpha_4 = - e^2\alpha_5 = \frac{s^2}{16}x^2 + O(x^4).
\end{equation}
The coefficients $\alpha_4$ and $\alpha_5$ provide the leading corrections to $WW$ and $WZ$ elastic scattering.  Note that the three-site model has $\alpha_2 \neq \alpha_3$ and therefore, $L_{9L} \neq L_{9R}$ \cite{Chivukula:2005ji,Falk:1991cm}.

%%%%%%%%%%%%%%%%%%%%%%%%%%%%%%%%%%%%
%%%%%%%%%%%%%%%%%%%%%%%%%%%%%%%%%%%%

\end{document}